\documentclass[aps,12pt]{revtex4}
\usepackage{epsfig}
\usepackage{graphicx}
\usepackage{amsmath,amssymb,color}
\usepackage[english]{babel}
\usepackage{latexsym}
\usepackage{amsfonts}
\usepackage{ulem}
\usepackage{algorithm, algorithmic}

\parskip=\medskipamount

\newcommand{\eq}[1]{(\ref{#1})}

\newcommand{\be}{\begin{equation}}
\newcommand{\ee}{\end{equation}}

\begin{document}

\title{Counting Phases and Faces Using  \\ Bayesian Thermodynamic Integration}

\author{Alexander Lobashev$^{1,* }$, Mikhail V. Tamm$^{2}$}

\affiliation{\bigskip
$^1$Skolkovo Institute of Science and Technology, Skolkovo, Russia \\
$^2$ERA Chair for Cultural Data Analytics, School of Digital Technologies, Tallinn University, Tallinn, Estonia}

\begin{abstract}

We introduce a new approach to reconstruction of the thermodynamic functions and phase boundaries in two-parametric statistical mechanics systems. Our method is based on expressing the Fisher metric in terms of the posterior distributions over a space of external parameters and approximating the metric field by a Hessian of a convex function. We use the proposed approach to accurately reconstruct the partition functions and phase diagrams of the Ising model and the exactly solvable non-equilibrium TASEP without any a priori knowledge about microscopic rules of the models. We also demonstrate how our approach can be used to visualize the latent space of StyleGAN models and evaluate the variability of the generated images.

\bigskip
$^*$ e-mail: Alexander.Lobashev@skoltech.ru

\end{abstract}

\maketitle

While machine learning methods are effective at analyzing statistical physics systems their application has mainly been limited to determining the boundaries of phase transitions and extracting the learned order parameters. However, in many cases it remains unclear how to relate these learned order parameters to known physical quantities that are commonly used to determine phase transitions.

We address here a general problem of reconstructing both phase boundaries and thermodynamic functions of a given statistical physics model. A model in this context is thought to be a stochastic mapping from a low-dimensional space of external parameters $\textbf{t}$ to a multi-dimensional space of microstates $s$. We don't assume any prior knowledge of the number of phases and their location in the space of external parameters. The only label which is available to us is the values of external parameters for which a given microstate was generated.

\textbf{Our contributions.} The contributions of this paper are as
follows:

\begin{itemize}
  \item We propose a new approach which we call Bayesian Thermodynamic Integration to approximate the Fisher information metric for the distributions over a high-dimensional microstate space depending on several external parameters. The approach is based on approximating the probability distribution with a function from the exponential family
  $$
  \mathbb{P}(s| \textbf{t}) = \frac{e^{f(s)^{T}\textbf{t}}}{Z( \textbf{t})}
  $$
  where the normalization factor (partition function in the terminology of statistical physics) is a convex function of parameters.
  For the statistical physics problems this approach gives access to the partition function and other thermodynamic functions without a priori knowledge of the system Hamiltonian;
  \item We apply the suggested approach to several two-parametric systems and demonstrate that it allows a satisfactory reproduction of the thermodynamic functions and phase transition lines. In the case of a two-parameter model of statistical mechanics based on the StyleGAN v3 generator, we observe signs of a second-order phase transition corresponding to the sharp changes of  identity of a generated human face. 
\end{itemize}

\section{Bayesian Thermodynamic Integration}
The main problem of interest to statistical physics is the study of phase transitions. They correspond to sharp changes in typical microstates with a gradual change in external parameters. We use Jensen-Shannon divergence to measure the distance between probability distributions corresponding to different values of external parameters
\begin{equation}
\begin{split}
    \text{JSD}(\mathbb{P}(s| \textbf{t}+\delta\textbf{t}), \mathbb{P}(s| \textbf{t})) =
    \mathcal{H}\left[\frac{1}{2}(\mathbb{P}(s| \textbf{t}+\delta\textbf{t})+ \mathbb{P}(s| \textbf{t}))\right] - \\ - \frac{1}{2}\left(\mathcal{H}[\mathbb{P}(s| \textbf{t}+\delta\textbf{t})]+ \mathcal{H}[\mathbb{P}(s| \textbf{t})]\right),
\end{split}
\end{equation}
where \text{JSD} is Jensen-Shannon divergence which is a proper metric on the manifold of probability distributions parameterized by $\textbf{t}$ and $\mathcal{H}[\mathbb{P}]$ represents entropy of the corresponding distribution. For small variations of $\textbf{t}$
\begin{equation}
    \text{JSD}(\mathbb{P}(s| \textbf{t}+\delta\textbf{t}), \mathbb{P}(s| \textbf{t})) = \delta\textbf{t}^{T}G(\textbf{t}) \delta\textbf{t} + O(\delta\textbf{t}^{3})
    \label{fisher}
\end{equation}
where G(\textbf{t}) is a Riemannian metric on the manifold of probability distributions called Fisher information metric \cite{amari2000methods}.

Fisher information metric is defined as the second order Taylor expansion term of KL divergence between close probability distributions
\begin{equation}
    G(\textbf{t}) = \int\mathbb{P}(s| \textbf{t}) \nabla_{\textbf{t}} \log\mathbb{P}(s| \textbf{t}) (\nabla_{\textbf{t}} \log \mathbb{P}(s| \textbf{t}))^{T} ds
    \label{gibbs_distribution}
\end{equation}
At phase transition points the Fisher metric is supposed to have singularities in the limit of large system size (i.e., for dimensionality of the space of $s$ approaching infinity).
Apply Bayes' formula to $\mathbb{P}(s| \textbf{t})$ to find posterior distribution on the space of external parameters given observed microstate
\begin{equation}
    \mathbb{P}(\textbf{t}|s) = \frac{\mathbb{P}(s|\textbf{t})\mathbb{P}(\textbf{t}) }{\mathbb{P}(s)}
    \label{bayes_formula}
\end{equation}
Thus, if prior distribution $\mathbb{P}(\textbf{t})$ is uniform then
\begin{equation}
    \nabla_{\textbf{t}} \log \mathbb{P}(\textbf{t}|s) = \nabla_{\textbf{t}} \log \mathbb{P}(s|\textbf{t})
    \label{rgularity_from_bayes_formula}
\end{equation}
This implies that, if one knows the posterior distribution $\mathbb{P}(\textbf{t}|s)$ (for example, via some sort of Monte-Carlo approximation), then the Fisher information metric can be expressed as
\begin{equation}
\begin{split}
    G(\textbf{t}) \approx \frac{1}{K}\sum_{k=1}^{K} \nabla_{\textbf{t}} \log\mathbb{P}(\textbf{t}|s_{k}) (\nabla_{\textbf{t}} \log \mathbb{P}(\textbf{t}|s_{k}))^{T}, \\ s_{k}\sim \mathbb{P}(s|\textbf{t})
\end{split}
    \label{gibbs_distribution}
\end{equation}

\subsection{Approximation of the Posterior}

Posterior distribution $\mathbb{P}(\textbf{t}|s)$ works as a sort of ``probabilistic thermometer": it looks at a microstate $s$ and gives a probabilistic forecast of external parameters (for example temperature) at which this microstate was generated.

By definition of the problem, it is possible to sample from $\mathbb{P}(s|\textbf{t})$, giving pairs 
$$(s_{i}, \textbf{t}_{i}) \sim \mathbb{P}(s|\textbf{t})\mathbb{P}(\textbf{t}),$$
sampled from joint distribution, where $\mathbb{P}(\textbf{t})$ is chosen to be uniform in some compact domain. These samples can be used to fit parametric family of distributions $\mathbb{P}_{\theta}(\textbf{t}|s)$ to approximate posterior  $\mathbb{P}(\textbf{t}|s)$. Here $\theta$ are the parameters of the family of distribution, which could be obtained by maximizing likelihood of true external parameters $\textbf{t}_{i}$ given $s_{i}$ over all our samples or equivalently by minimizing negative log-likelihood
\begin{equation}
    \text{NegativeLogLikelihood}(\theta) = -\sum_{i=1}^{N}\log(\mathbb{P}_{\theta}(\textbf{t}_{i}|s_{i}))
    \label{neg_log_likelihood}
\end{equation}
\begin{equation}
\end{equation}
Minimization of the log-likelihood \eq{neg_log_likelihood} is equivalent to minimization of the KL divergence between the distribution of target labels and the predicted distribution 
\begin{equation}
    \mathcal{L}(\theta) = \mathbb{E}_{\textbf{t}' \sim \mathbb{P}(\textbf{t})} \text{KL}(\mathbb{P}_{\text{target}}(\textbf{t}|\textbf{t}')|| \mathbb{P}_{\theta}(\textbf{t}|s(\textbf{t}'))),
\end{equation}
where $s(\textbf{t}')$ is a microstate of the physical model sampled from the conditional distribution $\mathbb{P}(s | \textbf{t}')$. If labels are considered hard, the target distribution is $ \mathbb{P}_{\text{target}}(\textbf{t}|\textbf{t}') = \delta (\textbf{t}-\textbf{t}')$. In order to avoid divergences, it is convenient to use a smoothened distribution 
\begin{equation}
    \mathbb{P}_{\text{target}} (\textbf{t}|\textbf{t}') = C \cdot \exp\left(-\frac{1}{2 \sigma^{2}}||\textbf{t}'-\textbf{t}||^{2}\right),
    \label{smoothed_target}
\end{equation}
instead. Here $C$ is the normalizing constant, it is obtained by integrating over the domain of interest in the space of macroscopic parameters. The standard deviation $\sigma$ is defined to be equal to the mean distance from a point $t$ in the training dataset to its K-th nearest neighbor.

The resulting parametric distribution can be used as input of formula (\ref{gibbs_distribution}) in order to get an approximation of the Fisher metric. However, this approach has two important drawbacks. 

First, in order to obtain a decent approximation the parametric family $\mathbb{P}_{\theta}$ should be flexible enough, implying a large number of parameters $\theta$. 
However, if $\dim(\theta) >> N$, i.e. if the number of parameters is much larger than the number of examples in the training set, the problem of overfitting arises.

Second, the Fisher metric depends on the derivative of the approximated function, and derivatives of approximated functions are typically much noisier than the functions themselves. As a result, without additional efforts to smooth the posterior distribution, approximation (\ref{gibbs_distribution}) turns out to be too noisy to be useful in practice.

In order to resolve the first problem we artificially augment the training set. Instead of using a single microstate $s$ as an input for the prediction of external parameters $\textbf{t}$ we use a randomly shuffled tuple of $K$ microstates $s_{1}, \dots, s_{K}$ generated at external parameters close to $\textbf{t}$. That is to say, start with a set of $N$ training pairs $(s_{i}, \textbf{t}_{i})$, for each $\textbf{t}_i$ choose $K$ values $\textbf{t}_j$ with the smallest distance $||\textbf{t}_i -\textbf{t}_j||$, form a randomly shuffled tuple out of the corresponding microstates, and use the set  $(\{s_{i,1},\dots, s_{i,K}\}, \textbf{t}_{i})$ as a training set, so that instead of $\mathbb{P}_{\theta}(\textbf{t}_{i}|s_{i})$ we are now training the model to predict $\mathbb{P}_{\theta}(\textbf{t}_{i}|s_{i, 1}, \dots, s_{i, K})$. This trick allows to increase the size of the training set by a factor $K!$ without generating any new microstates. The drawback is that the resolution with respect to $\textbf{t}$ decreases with growing $K$. However, this decrease is much slower, then the factorial growth of the size of the training set.

The resulting target distribution $\mathbb{P}_{\text{target}} (\textbf{t}|\textbf{t}')$ is obtained by averaging target distributions \eq{smoothed_target} over the $K$ nearest neighbours

\begin{equation}
    \mathbb{P}_{\text{target}}(\textbf{t}|\textbf{t}') = \frac{1}{K}\sum_{\hat{\textbf{t}}\in \text{Neib}(\textbf{t}')} C \cdot \exp\left(-\frac{1}{2 \sigma^{2}}||\hat{\textbf{t}}-\textbf{t}||^{2}\right),
    \label{final_smoothed_target}
\end{equation}

In order to resolve the second issue outlined above we approximate the posterior distribution with a distribution from the exponential family parametrized by convex function, as described in the next section.

\subsection{Free Energy Approximation via Convex Neural Network}

We know that Fisher metric is a positive-definite matrix $
\delta \textbf{t}^{T} G(\textbf{t}) \delta \textbf{t} \geq 0 \ \ \forall \ \delta \textbf{t}$. The main idea of our approach is to approximate this positive-definite matrix by the Hessian of a convex function.
Note that this means that if $\dim(\textbf{t})=1$, i.e. if there is a single scalar external macroscopic parameter, then this approximation is exact.

More precisely, we approximate the posterior distribution as a function of external parameters by a distribution from the so-called exponential family of distributions. It is instructive to introduce this family in a following way. Suppose for each value of $\textbf{t}$ we fix average values of some macroscopic observables $\int \mathbb{P}(s|\textbf{t}) f(s)ds = \Gamma(\textbf{t})$. We are interested in a probability distribution $\mathbb{P}(s)$, which maximizes entropy
\begin{equation}
    \mathcal{H}[\mathbb{P}(s)]= - \int \mathbb{P}(s) \log \mathbb{P}(s) ds
\end{equation}
while respecting this constraint. Maximizing the functional 
\begin{equation}
\begin{split}
    L[\mathbb{P}(s)] = \mathcal{H}[\mathbb{P}(s)]-  \textbf{t}^{T}\left(\int \mathbb{P}(s) f(s) ds-  \Gamma(\textbf{t})\right) - \\ -\lambda_{0} \left(\int \mathbb{P}(s)ds- 1\right),
\end{split}
\end{equation}
where $\textbf{t}$, and $\lambda_{0} $ are Lagrange multipliers, gives the exponential family distributions of the form
\begin{equation}
\mathbb{P}(s| \textbf{t}) = \frac{e^{f(s)^{T}\textbf{t}}}{Z( \textbf{t})} = \underset{\mathbb{P}(s)}{\text{argmax}} L[\mathbb{P}(s)]
\label{exponential-family}
\end{equation}
where the normalization factor
\begin{equation}
    Z( \textbf{t}) = \int e^{f(s)^{T}\textbf{t}} ds
\end{equation}
is known as ``the partition function'' in equilibrium statistical mechanics.


For distributions from exponential family Fisher metric reduces to Hessian of the logarithm of the partition function
\begin{equation}
\begin{split}
    G(\textbf{t}) = \int\mathbb{P}(s| \textbf{t}) \nabla_{\textbf{t}} \log\mathbb{P}(s| \textbf{t}) (\nabla_{\textbf{t}} \log \mathbb{P}(s| \textbf{t}))^{T} ds = \\ = \nabla_{\textbf{t}\textbf{t}}  \log Z( \textbf{t}).
\end{split}
\end{equation}
while its first derivatives are the mean values of the function $f(s)$ and are known as ``thermodynamic forces"  
\begin{equation}
\int f(s) \mathbb{P}(s| \textbf{t}) ds = \nabla_{\textbf{t}}  \log Z( \textbf{t})
\label{forces}
\end{equation}
Fisher metric tensor is known to be positive-definite, and thus the function $\log Z( \textbf{t})$ is convex.

Now, in order to solve the problem of approximating the derivatives of the posterior distribution $\mathbb{P} (\textbf{t}|s)$, we want to find a  function $Z_{w}(\textbf{t})$, depending on some set of parameters $w$ which is, on the one hand, convex, and on the other hand, such that the Jensen-Shannon divergence between the approximate posterior $\mathbb{P}_{\theta}(\textbf{t}|s)$ and the posterior distribution generated from the exponential family with partition function $Z_{w}(\textbf{t})$ is minimized.


If our prior is uniform then exponential-family posterior distribution takes the form
\begin{equation}
    \mathbb{P}_{w}^{(\text{eq})}(\textbf{t}| s) = \frac{\mathbb{P}_{w}(s|\textbf{t})\mathbb{P}(\textbf{t})}{\int \mathbb{P}_{w}(s|\textbf{t})\mathbb{P}(\textbf{t}) d \textbf{t}}= \frac{e^{f(s)^{T}\textbf{t}-\log Z_{w}( \textbf{t})}}{\int e^{f(s)^{T}\textbf{t}-\log Z_{w}( \textbf{t})}d \textbf{t}}
\end{equation}
Function $f(s)$ is unknown, but since we use a tuple of microstates to estimate a single posterior, we replace $f(s)$ by its expectation (\ref{forces}),
$$
\frac{1}{K}\sum_{k=1}^{K}f(s_{k}) \approx \nabla_{\textbf{t}}\log Z( \textbf{t}), \ s_{k} \sim P(s|\textbf{t}).
$$

To find parameters $w$ we minimize Jensen-Shannon divergence between posterior $\mathbb{P}_{\theta}(\textbf{t}|s)$ predicted using maximum likelihood and posterior $\mathbb{P}_{w}(\textbf{t}| s)$ with $f(s)$ replaced by the gradient of the partition function
\begin{equation}
P_{w}^{(\text{eq})}(\textbf{t}|s(\textbf{t}')) = \frac{\exp\left[(\nabla_{\textbf{t}'}\log Z_{w}( \textbf{t}'))^{T}\textbf{t}-\log Z_{w}( \textbf{t})\right]}{\int \exp\left[(\nabla_{\textbf{t}'}\log Z_{w}( \textbf{t}'))^{T}\textbf{t}-\log Z_{w}( \textbf{t})\right]d \textbf{t}}
\end{equation}
\begin{equation}
    L(w) = \mathbb{E}_{\textbf{t}' \sim \mathbb{P}(\textbf{t})}\text{JSD}\left(\mathbb{P}_{\theta}(\textbf{t}|s(\textbf{t}')), P_{w}^{(\text{eq})}(\textbf{t}|s(\textbf{t}')) \right)
\end{equation}
\begin{equation}
    w_{\text{optimal}} = \underset{w}{\text{argmin}}\ L(w)
\end{equation}
The resulting approximation for the Fisher metric is
$$
G(\textbf{t}) = \nabla_{\textbf{t}\textbf{t}} \log Z_{w_{\text{optimal}}}(\textbf{t}).
$$

We summarize the procedure described in this section by two algorithms outlined below, the first one generates the training dataset (\ref{alg:dataset-generation}), the second one finds the approximations for the partition function and the Fisher metric (\ref{alg:bayesian-thermodynamic-integration}).

\begin{algorithm}[th]
   \caption{Dataset Generation for Bayesian Thermodynamic Integration}
   \label{alg:dataset-generation}
\begin{algorithmic}
    \STATE {\bfseries Input:}\\ 
    \hspace{1em} $\mathbb{P}(s| \textbf{t})$ - likelihood function represented by an algorithm to sample data, \\ 
    \hspace{1em} $\mathbb{P}(\textbf{t})$ - uniform prior distribution on the space of external parameters, \\ 
    \hspace{1em} $N_{\text{dataset}}$ - dataset length, \\
    \hspace{1em} $\sigma^{2}$ - variance of the target gaussian distribution, \\
    \hspace{1em} $\mathcal{D} = \{\hat{\textbf{t}}_{1}, \dots, \hat{\textbf{t}}_{N_{\mathcal{D}}}\}$ - a discretization of the support of the uniform prior distribution $\mathbb{P}(\textbf{t})$. \\
    \hspace{1em} $\text{L}_{\text{sam}}$ = [] - list of input samples, \\ 
    \hspace{1em} $\text{L}_{\text{par}}$ = [] - list of external parameters, \\
    \hspace{1em} $\text{L}_{\text{tar}}$ = [] - list of target posterior distributions
    \FOR{$i=1$ {\bfseries to} $N_{\text{dataset}}$ }
    \STATE sample $\textbf{t}_{i} \sim \mathbb{P}(\textbf{t})$ and append $\textbf{t}_{i}$ to $\text{L}_{\text{par}}$ \\
    \STATE sample $s_{i} \sim \mathbb{P}(s| \textbf{t}_{i})$ and append $s_{i}$ to $\text{L}_{\text{sam}}$ \\
    $p_{\text{tar}}^{(i)}$ = [] - discretized target probability distributions
    \FOR{$j=1$ {\bfseries to} $N_{\mathcal{D}}$ }
    \STATE  $
    p_{\text{tar}}^{(i)}[j] = 
    \exp\left(-\frac{1}{2 \sigma^{2}}||\hat{\textbf{t}}_{j}-\textbf{t}_{i}||^{2}\right)
    $
    \ENDFOR
    \STATE $p_{\text{tar}}^{(i)} = \text{normalize}(p_{\text{tar}}^{(i)})$ and append $p_{\text{tar}}^{(i)}$ to $\text{L}_{\text{tar}}$
    \ENDFOR
    \STATE {\bfseries Output:} \ 
    \\
    \hspace{1em} $\text{L}_{\text{sam}}$ - dataset of input samples, \\ 
    \hspace{1em} $\text{L}_{\text{par}}$ - dataset of external parameters, \\
    \hspace{1em} $\text{L}_{\text{tar}}$ - dataset of target posterior distributions
\end{algorithmic}
\end{algorithm}

\begin{algorithm}[th]
   \caption{Bayesian Thermodynamic Integration}
   \label{alg:bayesian-thermodynamic-integration}
\begin{algorithmic}
    \STATE {\bfseries Input:} \\  \hspace{1em} $\text{L}_{\text{sam}}$ - train dataset of input samples, \\ 
    \hspace{1em} $\text{L}_{\text{par}}$ - train dataset of external parameters, \\
    \hspace{1em} $\text{L}_{\text{tar}}$ - train dataset of target posterior distributions, \\
    \hspace{1em} $K_{\text{bundle}}$ - bundle size or the number of nearest neighbours, \\
    \hspace{1em} $N_{\text{U2Net steps}}$ - steps for posterior approximation. \\
    \hspace{1em} $N_{\text{ICNN steps}}$ - steps for free energy approximation.\\
    \hspace{1em} $\mathbb{P}_{\theta}(\textbf{t}|s_{1}, \dots, s_{K_{\text{bundle}}})$ - neural network with output shape equal to the shape of elements of $\text{L}_{\text{tar}}$, \\
    \hspace{1em} $\text{L}_{\text{neib}}$ = [] - list of $K_{\text{bundle}}$ nearest neighbors of points in $\text{L}_{\text{par}}$. \\
    \STATE $\text{L}_{\text{neib}} = \text{KNN}(\text{L}_{\text{par}}, K_{\text{bundle}}, \text{include\_self=True})$
    \FOR{$i=1$ {\bfseries to} $N_{\text{U2Net steps}}$}
    \STATE choose $\textbf{t}'$ randomly from $\text{L}_{\text{par}}$
    \STATE $s(\textbf{t}') = (s_{1}, \dots, s_{K_{\text{bundle}}})$ shuffle and concatenate $K_{\text{bundle}}$ microstates corresponding to the nearest neighbors of $\textbf{t}'$ taken from $\text{L}_{\text{neib}}$
    \STATE $\mathbb{P}_{\text{target}} = \frac{1}{K_{\text{bundle}}}\sum_{\hat{\textbf{t}}\in \text{Neib}(\textbf{t}')} \text{L}_{\text{tar}}[\text{index}(\hat{\textbf{t}})]$ 
    \STATE $\mathcal{L}(\theta) = \mathbb{E}_{\textbf{t}' \sim \mathbb{P}(\textbf{t})} \text{KL}(\mathbb{P}_{\text{target}}(\textbf{t}|\textbf{t}')|| \mathbb{P}_{\theta}(\textbf{t}|s(\textbf{t}')))$
    \STATE $\theta \leftarrow \text{Adam}(\theta, \nabla_{\theta} \mathcal{L}(\theta))$
    \ENDFOR \\
    \hspace{1em} $Z_{w}(\textbf{t})$ - input convex neural network,
    \FOR{$i=1$ {\bfseries to} $N_{\text{ICNN steps}}$}
    \STATE $\mathbb{P}_{w}^{(\text{eq})}(\textbf{t}|\textbf{t}') = \frac{e^{(\nabla_{\textbf{t}'}\log Z_{w}( \textbf{t}'))^{T}\textbf{t}-\log Z_{w}( \textbf{t})}}{\int e^{(\nabla_{\textbf{t}'}\log Z_{w}( \textbf{t}'))^{T}\textbf{t}-\log Z_{w}( \textbf{t})}d \textbf{t}}$
    \STATE $\mathcal{L}(w) = \mathbb{E}_{\textbf{t}' \sim \mathbb{P}(\textbf{t})}\text{JSD}\left(\mathbb{P}_{\theta}(\textbf{t}|s(\textbf{t}')), \mathbb{P}_{w}^{(\text{eq})}(\textbf{t}|\textbf{t}') \right)$
    \STATE $w \leftarrow \text{Adam}(w, \nabla_{w} \mathcal{L}(w))$
    \ENDFOR
    \STATE {\bfseries Output:} \\
    \hspace{1em} $F(\textbf{t}) = \log Z_{w}(\textbf{t})$ - approximated free energy, \\
    \hspace{1em} $G(\textbf{t}) = \nabla_{\textbf{t}\textbf{t}} \log Z_{w}(\textbf{t})$ - approximated Fisher metric. \\
    \ \ \\
\small{
KNN = k-nearest neighbors \\
KL = Kullback–Leibler divergence \\
JSD = Jensen–Shannon divergence
}
\end{algorithmic}
\end{algorithm}

\section{Numerical experiments}

\subsection{Ising model}

\begin{figure}[ht]
\vskip 0.2in
\begin{center}
\centerline{\includegraphics[width=\columnwidth]{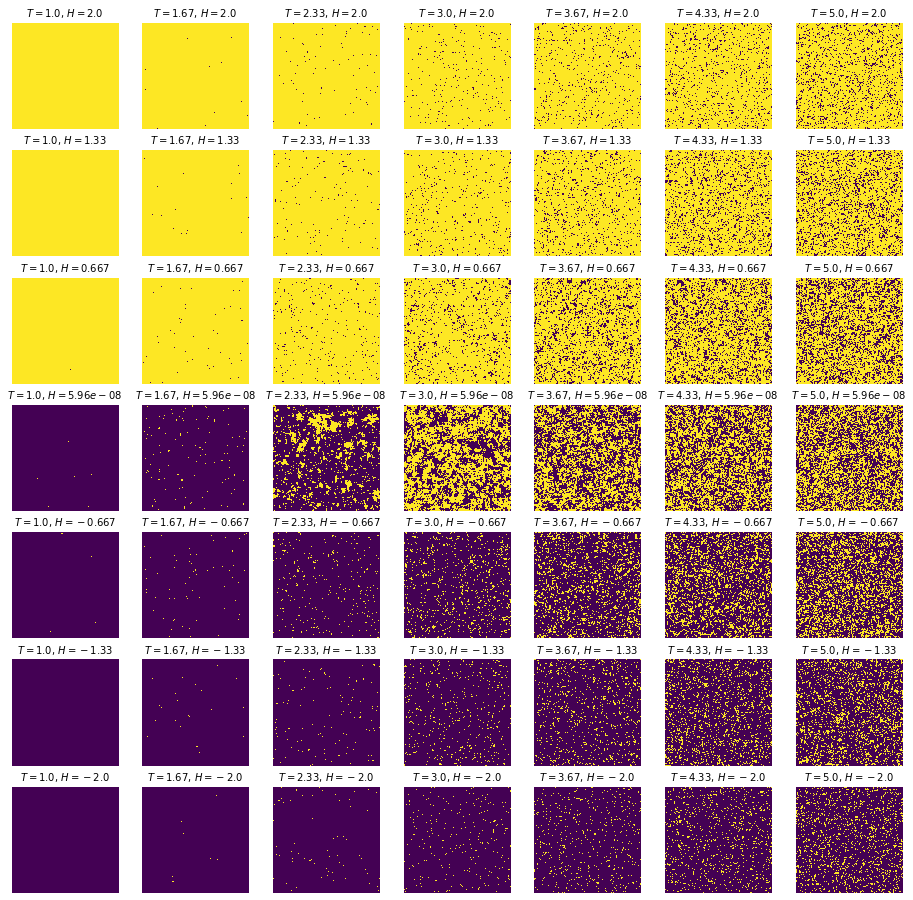}}
\caption{Examples of the microstates of the 2D Ising model, temperature increases from right to left, magnetic field increases from bottom to top.}
\label{ising-states-grid}
\end{center}
\vskip -0.2in
\end{figure}

In this section we test whether our approach is capable of reconstructing thermodynamic functions for an equilibrium statistical mechanics model in which the distribution of microstates belongs to the exponential family. We consider a 2D Ising model \cite{ising1925beitrag}, which is an archetypal model of phase transitions in statistical mechanics. A microstate of this model is a set $s$ of spin variables $s_i = \pm 1$ defined on each vertex of a square lattice of size $L\times L$. At equilibrium probability distribution over the space of microstates is
\begin{equation}
    \mathbb{P}(s|H, T) = \frac{1}{Z(H, T)} e^{- \beta \sum_{\langle i,j \rangle}s_{i}s_{j} - \beta H \sum_{i} s_{i}}
    \label{ising_likelihood}
\end{equation}
where $H$ and $T=1/\beta$ are external parameters called magnetic field and temperature, respectively, the first sum is over all edges of the lattice, and $Z(H,T)$ is a normalization parameter known as a partition function:
\begin{equation}
    Z(H, T)= \sum_{s_1=\pm1}\dots\sum_{s_N=\pm1} e^{- \beta \sum_{\langle i,j \rangle}s_{i}s_{j} - \beta H \sum_{i} s_{i}}.
\end{equation}
This model is exactly solvable for $H=0$ \cite{onsager1944crystal, kac1952combinatorial, baxter1978399th}. In particular, it is known that at $T_{\text{cr}} = \frac{2}{\log (1+\sqrt{2} )} \approx 2.269$ a transition occurs between the high-temperature disordered state, where spin variables are on average equal zero, and the low-temperature ordered state in which average value of spin becomes distinct from zero. For general values of $H \neq 0$ the likelihood function \eq{ising_likelihood} is intractable.

Our dataset consists of $N=540000$ samples of spin configurations on the square lattice of size $L\times L =128 \times 128$ with periodic boundary conditions. We consider the parameter ranges $T \in [T_{\min}, T_{\max}]=[1, 5], H \in [H_{\min}, H_{\max}]=[-2, 2]$ similar to the ranges used in (Walker, 2019). Point $(T, H)$ is sampled uniformly from this rectangle, and then a sample spin configuration is created for these values of temperature and external field by starting with a random initial condition and equilibrating is with Glauber (one-spin Metropolis) dynamics for $10^{4}\times 128 \times 128 \approx 1.64 \times 10^{8}$ iterations. We represent spin configuration as a single-channel image with values $+1$ and $-1$. When constructing target probability distributions we choose $\sigma=\frac{1}{50}$ and set the discretization $\mathcal{D}$ of the square $[T_{\min}, T_{\max}]\times[H_{\min}, H_{\max}] = [1, 5]\times[-2, 2]$ to be a uniform grid with $L\times L =128 \times 128$ grid cells.

\begin{figure}[ht]
\vskip 0.2in
\begin{center}
\centerline{\includegraphics[width=\columnwidth]{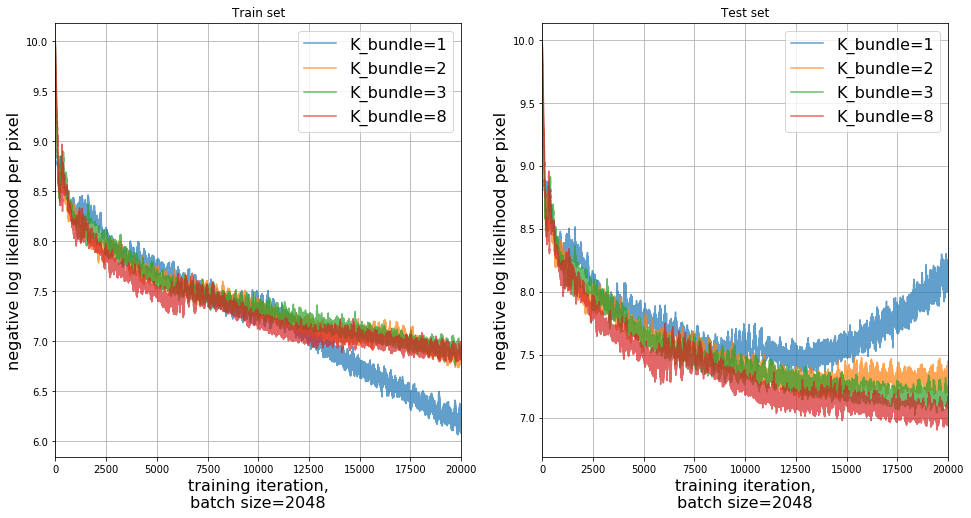}}
\caption{2D Ising model. Train (left) and test (right) loss dynamics for different values of $K_{\text{bundle}}$ in the bundle augmentation procedure.}
\label{bundle-augmentation}
\end{center}
\vskip -0.2in
\end{figure}

\par Image-to-image network with U$^{2}$-Net architecture \cite{qin2020u2} is used to approximate posterior $\mathbb{P}_{\theta}(\textbf{t}|s)$. The network takes as input a bundle of $K_{\text{bundle}}$ images concatenated across channel dimension and outputs a categorical distribution representing density values in discrete grid points. For simplicity we choose the discretization $\mathcal{D}$ to be of the same spatial dimensions as the input image. For all our numerical experiments the training was performed on a single Nvidia-HGX compute node with 8 A100 GPUs. We trained U$^{2}$-Net using Adam optimizer with learning rate $0.00001$ and batch size of $2048$ for $N_{\text{U2Net steps}}=20000$ gradient update steps. In all our experiments the training set consists of 80\% of samples and the other 20\% are used for testing.

\textbf{Bundle augmentation.} To explore how test loss depends on the bundle size $K_{\text{bundle}}$, which determines the number of microstates used to evaluate a single posterior distribution, we performed four different training runs with $K_{\text{bundle}} \in \{1,2,3,8\}$. The results are shown in \ref{bundle-augmentation}. For $K_{\text{bundle}}=1$, we observed overfitting after about the 12000th training iteration. However, even with $K_{\text{bundle}}=2$, overfitting is significantly reduced and continues to decrease at large values of $K_{\text{bundle}}$. The downside of this approach is that the effective grid resolution of the posterior distribution decreases with growing $K_{\text{bundle}}$ as
$$
L_{\text{eff}} \approx \sqrt{\frac{N}{K_{\text{bundle}}}},
$$
where $N$ is the dataset size. In what follows, we set $K_{\text{bundle}}=4$. Also, since we are using a fully convolutional architecture, increasing the number of input channels by $K_{\text{bundle}}$ times has almost no effect on the overall GPU memory consumption during training.

\begin{figure}[ht]
\vskip 0.2in
\begin{center}
\centerline{\includegraphics[width=\columnwidth]{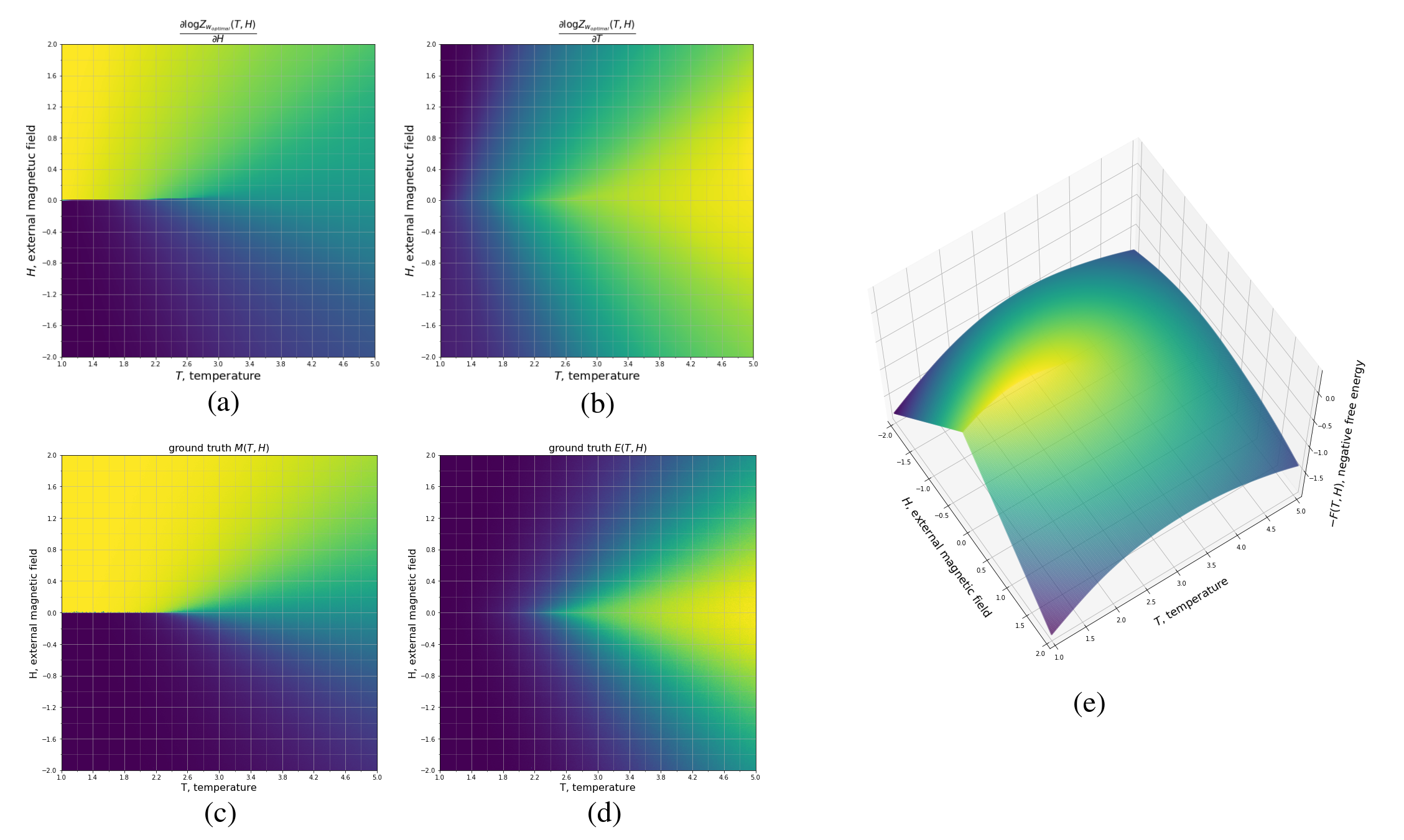}}
\caption{2D Ising model. (a) Partial derivative of the reconstructed free energy with respect to temperature $\frac{\partial F_{\text{rec}}(T, H)}{\partial T}$, (b) Partial derivative of the reconstructed free energy with respect to magnetic field $\frac{\partial F_{\text{rec}}(T, H)}{\partial H}$, (c) energy of the Ising model $E(H, T) = \sum_{\langle i,j \rangle}s_{i}(H, T)s_{j}(H, T)$, (d) magnetization of the Ising model $M(H, T) = \sum_{i} s_{i}$, (e) reconstructed free energy.}
\label{ising-free-energy-results}
\end{center}
\vskip -0.2in
\end{figure}

\textbf{Free energy approximation.}
We use Input Convex Neural Network \cite{amos2017input} architecture to approximate the free energy. The fully connected ICNN has 5 layers with hidden dimension 512. We train network using Adam optimizer with learning rate 0.001. 

Reconstructed free energy is shown in \ref{ising-free-energy-results}. The model was able to correctly determine 1st order phase transition line $T<T_{\text{cr}}$ and $H=0$.

\subsection{Totally asymmetric simple exclusion process}

\begin{figure}[ht]
\vskip 0.2in
\begin{center}
\centerline{\includegraphics[width=\columnwidth]{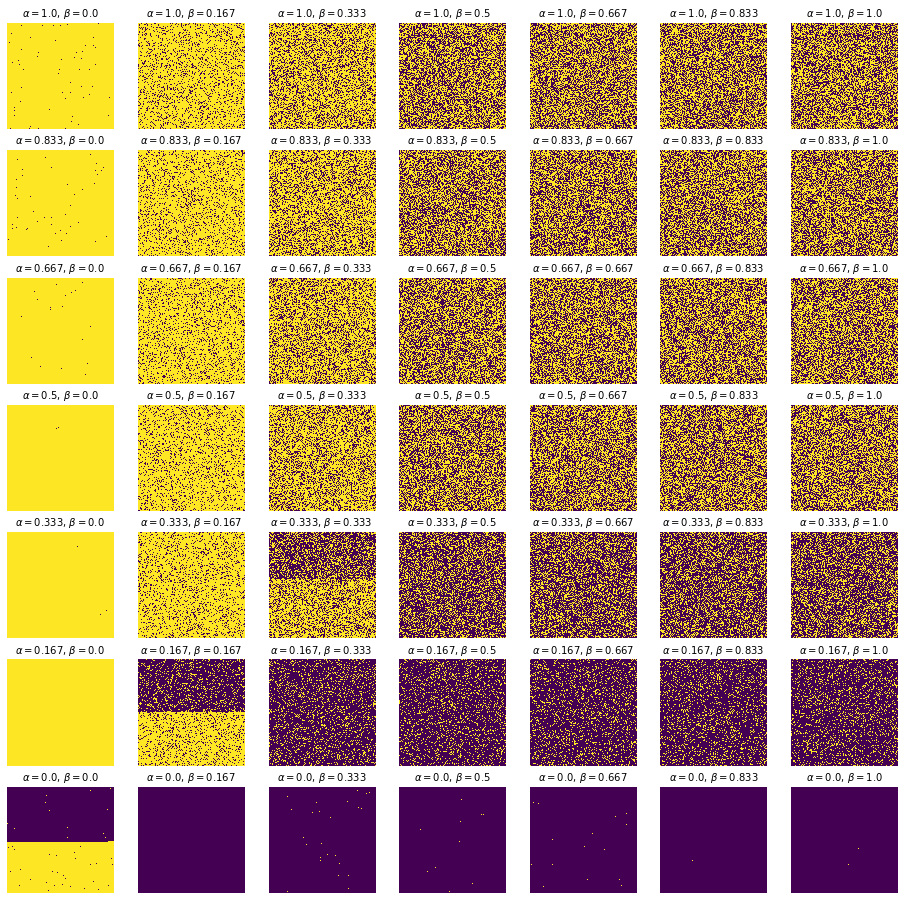}}
\caption{Examples of microstates of the stationary state TASEP. The microstate is presented in raster ordering, the cell numbers increase downwards from row to row, and rightwards within a row. $\alpha$ and $\beta$ increase from bottom to top and from left to right, respectively. Once can clearly see the high density (left), low-density (bottom) and maximal current (density = 1/2, top right) phases.}
\label{ising-states-grid}
\end{center}
\vskip -0.2in
\end{figure}

Totally asymmetric simple exclusion process (TASEP) is a simple model of 1-dimensional transport phenomena. A microscopic configuration is a set of particles on a 1d lattice respecting the condition that there can be no more than one particle in each lattice cell. Each particle can move to the site to the right of it with probability $p dt$ per time $dt$ provided that it is empty (we put $p=1$ without loss of generality). When complemented with boundary conditions, the TASEP attains a stationary state as time goes to infinity. 

One particular case is open boundary conditions, when a particle is added with probability $\alpha dt$ per time $dt$ to the leftmost site provided that it is empty and removed with probability $\beta dt$ per time $dt$ from the rightmost site provided that it is occupied. For this boundary condition the probability distribution is known exactly \cite{dehp, evans_review, krapivsky_book} and it once again takes the form
\begin{equation}
\mathbb{P}(s|\alpha, \beta) = \frac{f(s|\alpha, \beta)}{Z(\alpha, \beta)} ;\;\;     Z(\alpha, \beta)= \sum_{s}f(s|\alpha, \beta);
\end{equation}
where microstate $s$ is a concrete sequence of filled and empty cells, and $f(s|\alpha, \beta)$ is some function of $s$ and external parameters $\alpha, \beta$. Importantly, however, the function $f$, which is known exactly for all system sizes $M$ and all values of $s,\alpha, \beta$  does not take the form (\ref{exponential-family}). TASEP with free boundaries exhibits a rich phase behavior: for large system sizes three distinct phases - the low-density phase, the high-density phase and the maximal current phase are possible depending on the values of $\alpha, \beta$, and the asymptotic "free energy" which is defined as
\begin{equation}
F_{\text{TASEP}}(\alpha, \beta) = \lim_{M\to\infty} M^{-1} \log  Z(\alpha, \beta)
\end{equation}
and coincides with average flow per unit time, equals
\begin{equation}
F_{\text{TASEP}}(\alpha, \beta) = 
\begin{cases}
\frac{1}{4}, \ \alpha>\frac{1}{2}, \ \beta>\frac{1}{2};\\
\alpha(1-\alpha), 
 \ \alpha<\beta, \ \alpha<\frac{1}{2};\\
\beta(1-\beta), 
\ \beta<\alpha, \ \beta<\frac{1}{2}.\\
\end{cases}
\end{equation}
where the first, second and third cases correspond to maximal current, high density and low density cases, respectively.

We generate a dataset of $N=150000$ stationary TASEP configurations on a 1d lattice with $M=16384$ sites. The rates $\alpha (\beta)$ of adding (removing) particles at the left(right) boundary are sampled from the uniform prior distribution over a square $[0,1] \times [0,1]$. To reach the stationary state we start from a random initial condition and perform $N_{\text{steps}} = 2\times 10^{9} \approx 8 M^2$ move attempts, which is known to be enough to achieve the stationary state except for the narrow vicinity of the transition line $\alpha = \beta<1/2$ between high-density and low-density phases (in this case the stationary state has a slowly diffusing front of a shock wave in it, one needs of order $M^2$ move attempts to form the shock but of order $M^3$ move attempts for it to diffusively explore all possible positions of the shock).

We reshape 1d lattice with $M=16384$ sites into an image of size $L\times L =128 \times 128$ using raster scan ordering. To construct target probability distributions we set $\sigma=\frac{1}{150}$ and define the discretization $\mathcal{D}$ as a uniform grid on $[\alpha_{\min}, \alpha_{\max}]\times[\beta_{\min}, \beta_{\max}] = [0,1]\times[0,1]$ with $L\times L =128 \times 128$ grid cells.

\begin{figure}[ht]
\vskip 0.2in
\begin{center}
\centerline{\includegraphics[width=\columnwidth]{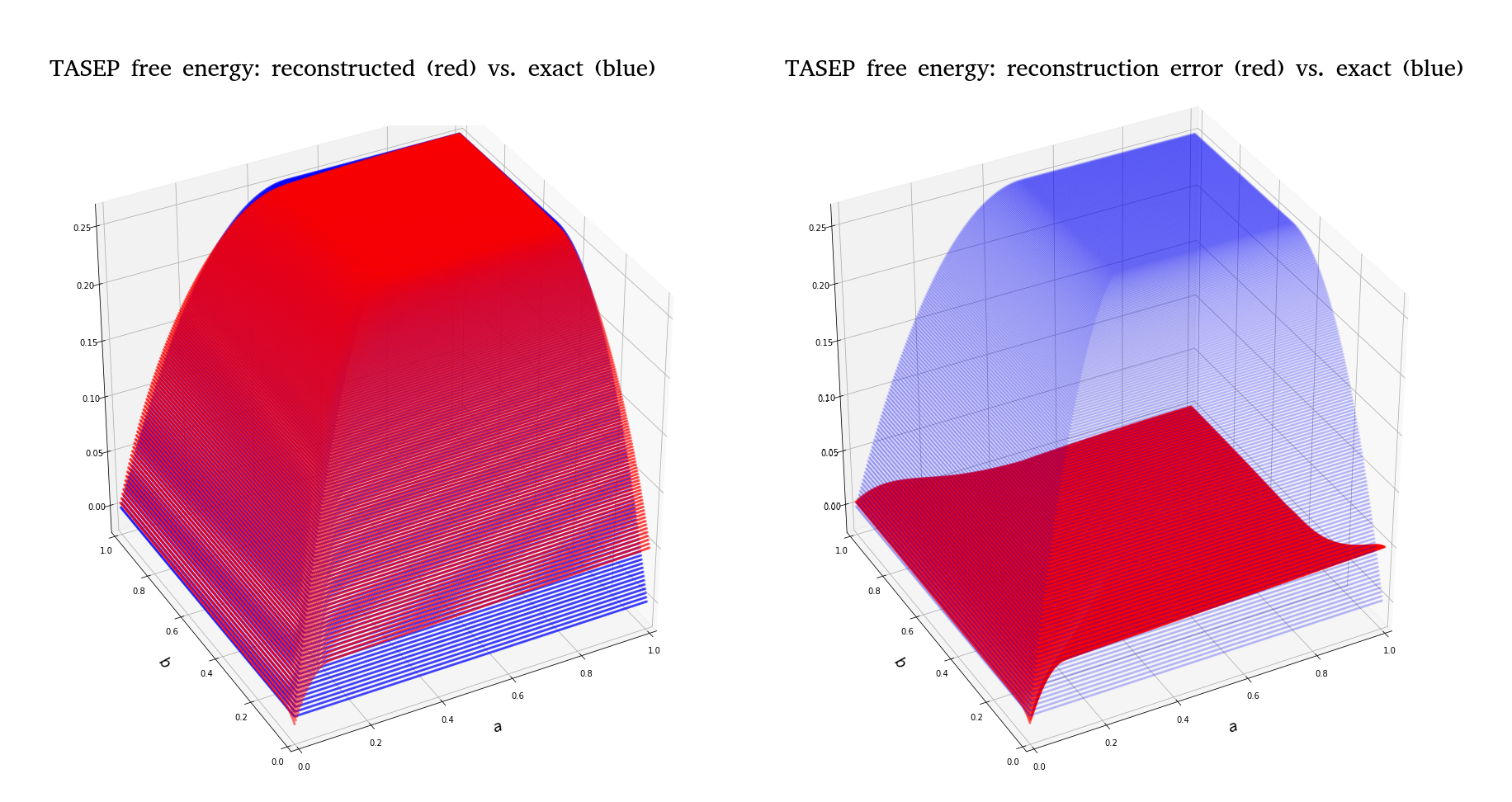}}
\caption{TASEP. Left: reconstructed free energy (red) compared to the exact solution (blue), right: reconstruction error (red) vs. exact solution.}
\label{tasep-free-energy-reconst}
\end{center}
\vskip -0.2in
\end{figure}

The comparison of the reconstructed free energy and the exact analytical solution is shown in \ref{tasep-free-energy-reconst}. 

\subsection{Image space of the StyleGAN}

\begin{figure}[ht]
\vskip 0.2in
\begin{center}
\centerline{\includegraphics[width=\columnwidth]{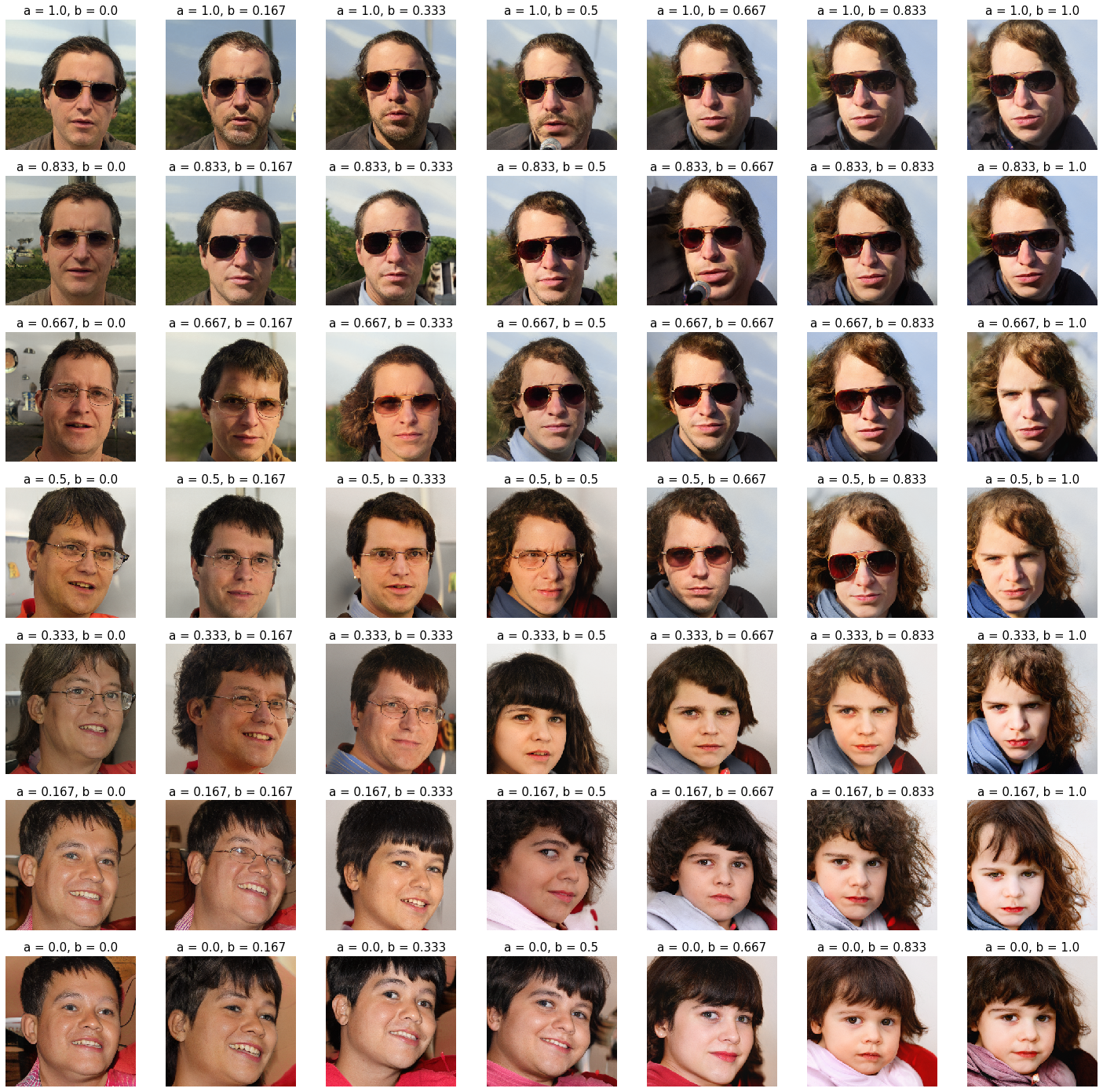}}
\caption{StyleGAN v3 pretrained on FFHQ dataset. Landscape of generated  images.}
\label{styleganv3-image-states-grid}
\end{center}
\vskip -0.2in
\end{figure}


Consider a synthetic two-parametric statistical mechanics model where microstates are sampled using the StyleGAN v3 generator \cite{karras2021alias}. Let $z_{1}, z_{2}$ and $z_{3}$ be vectors from the latent space of dimension 512. Consider a 2d section of the latent space parameterized by $a$ and $b$ as follows
\begin{equation}
    z(a,b) = z_{1}+a(z_{2}-z_{1})+b(z_{3}-z_{1}) + \xi,
\end{equation}
where $\xi$ is a normally distributed random vector with zero mean and small enough standard deviation (we it to be 1/5 of the standard deviation of the prior standard normal distribution). External parameters $a$ and $b$ lie in the rectangle $[0,1] \times [0,1]$, so that $z(a,b)$ interpolates between three base latent vectors at the corners $z(0,0) = z_{1}$, $z(1,0) = z_{2}$, $z(0,1) = z_{3}$ of the rectangle.
By putting  latent code $z(a,b)$ into the generator of StyleGAN we can sample images as functions of two external parameters.
\begin{equation}
    s = G(z(a,b))
\end{equation}

For FFHQ dataset we generate dataset of $N=500000$ human faces using StyleGAN at resolution of $1024 \times 1024$ and resize it to the resolution $128 \times 128$ before feeding it to the U2Net model which approximates posterior distribution on the space of external parameters. Experiment setup was similar to the 2D Ising model.

\begin{figure}[ht]
\vskip 0.2in
\begin{center}
\centerline{\includegraphics[width=\columnwidth]{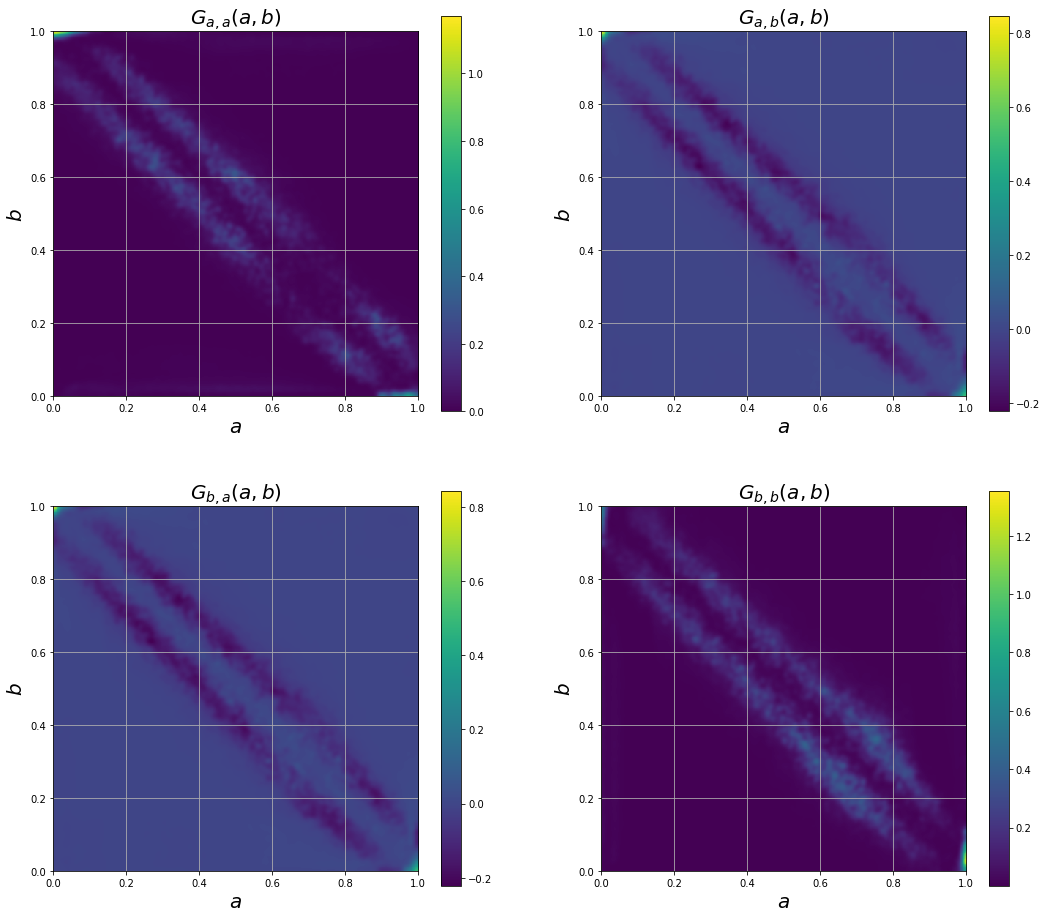}}
\caption{StyleGAN v3 with images as microstates.  Components of the Fisher information metric $G(a,b)$. (top left) $G_{aa}(a,b)= \frac{\partial^{2} \log Z(a, b)}{\partial a^{2}}$ component, (top right) $G_{ab}(a,b)== \frac{\partial^{2} \log Z(a, b)}{\partial a \partial b}$ component, (bottom left) $G_{ba}(a,b)=G_{ab}(a,b)$ component, (bottom right) $G_{bb}(a,b)= \frac{\partial^{2} \log Z(a, b)}{\partial b^{2}}$ component. Unexpectedly, we found two diagonal lines in the components of the Fisher metric, which, from the point of view of statistical mechanics, are signs of second-order phase transitions. }
\label{stylegan-images-fisher}
\end{center}
\vskip -0.2in
\end{figure}

Reconstructed Fisher metric is shown in \ref{stylegan-images-fisher}. Unexpectedly, we found two diagonal lines in the components of the Fisher metric, which, from the point of view of statistical mechanics, are signs of second-order phase transitions. If we compare it with the faces presented in \ref{styleganv3-image-states-grid} and \ref{generrated-faces-landscape-large} we can observe that phase transition lines correspond to the sharp identity change of generated faces. In the next section we investigate whether this behaviour is determined by the mapping network part of StyleGAN or it depends only on the synthesis network.

\subsection{Intermediate latent space $\mathcal{W}$ of the StyleGAN}

\begin{figure}[ht]
\vskip 0.2in
\begin{center}
\centerline{\includegraphics[width=\columnwidth]{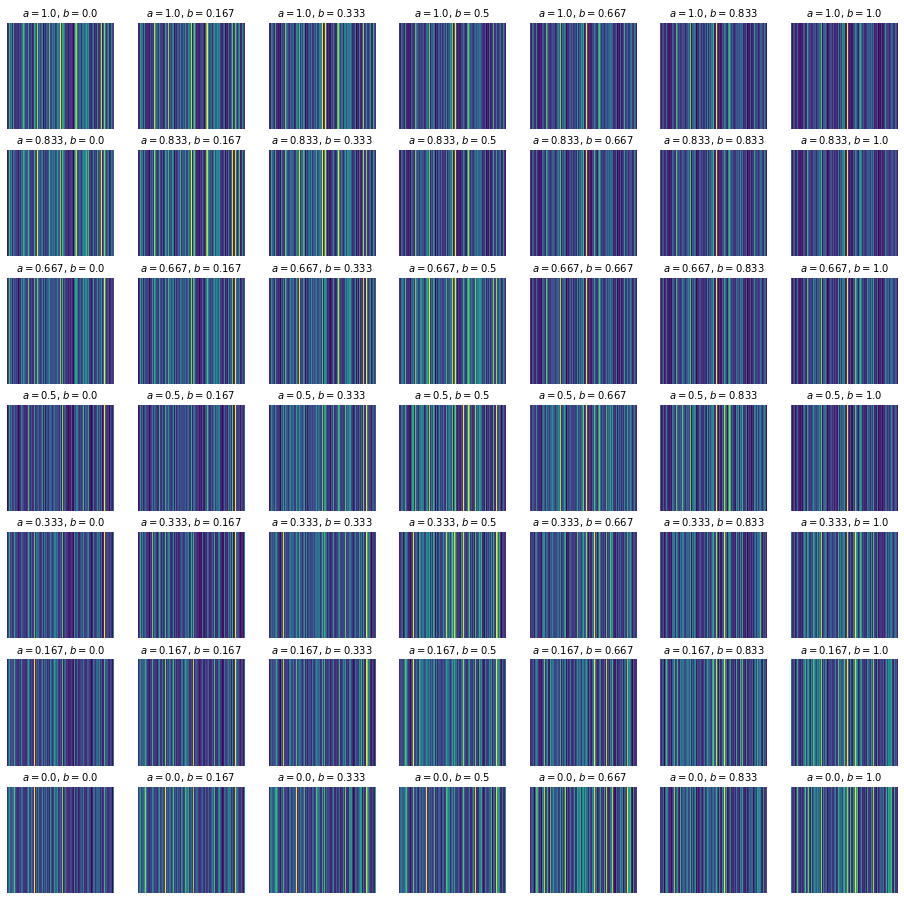}}
\caption{StyleGAN v3 pretrained on FFHQ dataset. Landscape of intermediate latent codes from the $\mathcal{W}$ space.}
\label{styleganv3-latent-states-grid}
\end{center}
\vskip -0.2in
\end{figure}

Given a latent vector $z$ the StyleGAN v3 generates images in two steps. First, the mapping network is applied to generate a vector (code) in the intermediate latent space $\mathcal{W}$. This latent code in case of the StyleGAN v3 model pretrained on FFHQ dataset is a matrix with shape $16\times512$ \cite{karras2021alias, karras2020analyzing}. Second, this vector is used by the weight demodulation layers of the synthesis network to generate the image itself.

A lot of attention was devoted to the improvement of the synthesis network. The effect of the mapping network on the generated images attracted less attention. 
To study which network, mapping or synthesis, is responsible for the typical shape of the Fisher metric field of the StyleGAN we introduce another synthetic two-parametric statistical mechanics model where microstates are generated by the mapping network

\begin{equation}
    w = G_{\text{mapping}}(z(a,b))
\end{equation}

Similarly to the previous section we generate $N=500000$ intermediate latent code of the shape $16 \times 512$ for the same latent vectors $z(a,b)$ and resize it into an image of size $128 \times 128$.

\begin{figure}[ht]
\vskip 0.2in
\begin{center}
\centerline{\includegraphics[width=\columnwidth]{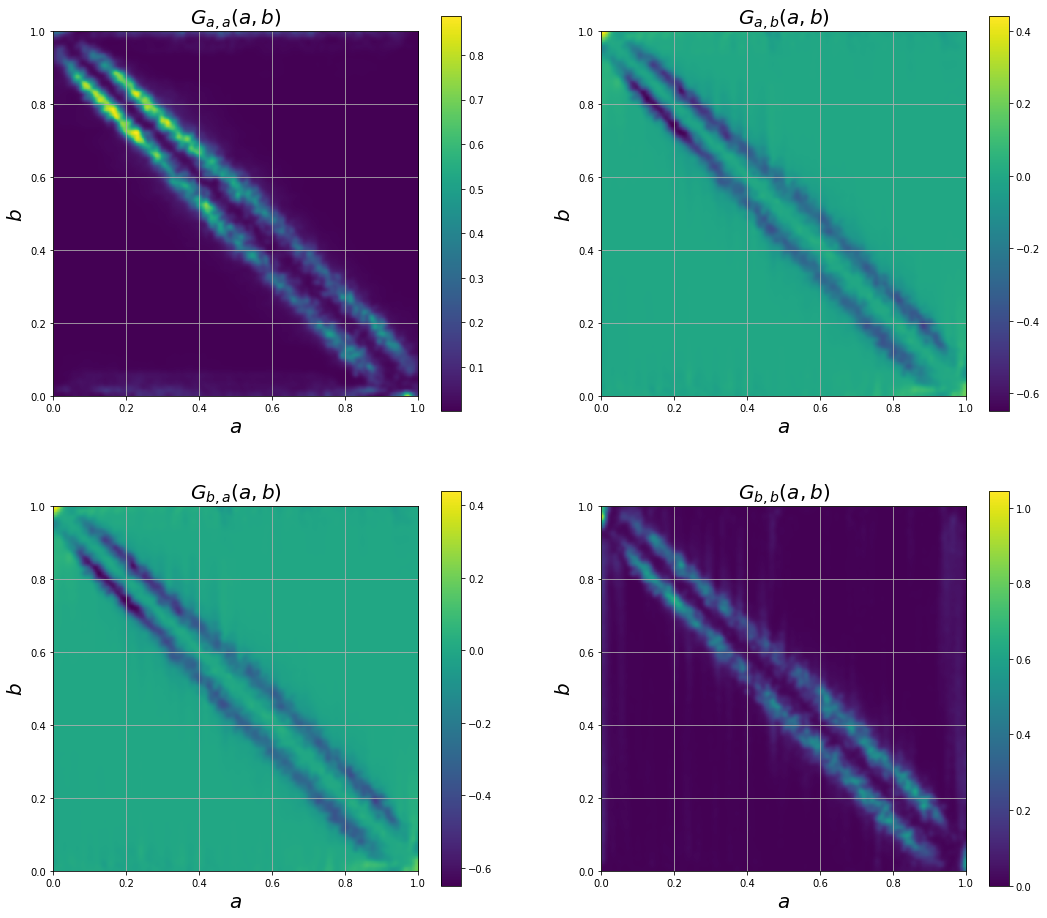}}
\caption{StyleGAN v3 with latent vectors from the  $\mathcal{W}$ space as microstates.  Components of the Fisher information metric $G(a,b)$. (top left) $G_{aa}(a,b)$ component, (top right) $G_{ab}(a,b)$ component, (bottom left) $G_{ba}(a,b)=G_{ab}(a,b)$ component, (bottom right) $G_{bb}(a,b)$ component. Compared to the components of the Fisher metric obtained in the previous section, the two lines are closer to each other, but their general arrangement remains the same.}
\label{stylegan-latent-fisher}
\end{center}
\vskip -0.2in
\end{figure}

As it can be seen from the \ref{stylegan-latent-fisher} and \ref{stylegan-images-fisher} Fisher metric fields for image space and for the $\mathcal{W}$ space almost coincide which means that mostly the mapping network defines variability of the generated faces and the synthesis network doesn't change the distribution of output images. In other words the synthesis network performs almost a bijective transformation of vectors from the intermediate latent space $\mathcal{W}$.

\section{Previous work}

Machine learning methods have been actively used in the study of models of classical and quantum statistical physics \cite{van2017learning, carrasquilla2017machine, rem2019identifying, wang2021unsupervised, canabarro2019unveiling}. The main problem of concern was to determine boundaries of phase transitions and to extract learned "order parameters" which can be used to distinguish one phase from another. Since equilibrium distribution of microstates of a statistical mechanics system belongs to the exponential family (usually with intractable normalization constant) the problem of determining "order parameters" is equivalent to the search of the sufficient statistics \cite{jiang2017learning, chen2020neural}. In \cite{walker2020deep} it was observed that principal components of the mean and standard deviation vectors learned by a variational autoencoder trained on the configurations of a 2D Ising model were highly correlated with the known physical quantities, indicating that variational autoencoder implicitly extract sufficient statistics from the data.

In contrast to \cite{chen2020neural} and \cite{walker2020deep} where neural networks were used to extract sufficient statistics from the raw microstates our approach is based on approximating the partition function directly and expressing mean values of the sufficient statistics through the derivatives of the logarithm of the partition function.

Our approach to the visualization of the StyleGAN latent space in terms of the Fisher metric could be thought as an unsupervised alternative to the approach used in \cite{tran2018dist} to study the GANs mode collapse problem, where the authors applied a classification network trained on MNIST to classify the generated images and construct a "phase diagrams” of predicted classes in the space of two-dimensional latent parameters. Similar approach was used in \cite{carrasquilla2017machine} to extract phase diagram of the Ising model by training network to classify low-temperature and high-temperature phases and drawing prediction on the space of external parameters. In our method, the singularities of the Fisher metric automatically highlight the boundaries of second-order phase transitions without an additional classification network or any other a priori knowledge about the number of phases and their location on the parameter space. However, we assume that all integrals over the space of external parameters are tractable, which is not the case for the general $512$-dimensional hidden space of StyleGAN.  Since $512$-dimensional integrals cannot be computed efficiently using domain discretization, other approaches are needed to parameterize the predicted posterior distributions on the parameter space, such as  tensor-train density estimation method \cite{novikov2021tensor}, which admits a tractable partition function.


\section{Discussion}

We propose a new approach to the reconstruction of the thermodynamic functions: partition function, free energy and their derivatives as functions of the external parameters, and apply it to several two-parametric statistical mechanics models. Our method is based on expressing the Fisher metric on the manifold of probability distributions over a high-dimensional space of microstates through the posterior distributions over a space of external parameters. Log-partition function is obtained by approximating the metric field by a Hessian of a convex function parametrized by an Input Convex Neural Network (ICNN).

The proposed approach is used to reconstruct the partition functions and phase diagrams of the equilibrium Ising model and the exactly solvable non-equilibrium TASEP without any a priori knowledge about microscopic rules of the models. The only information we need is some algorithm allowing to sample microstates for given values of the external parameters.

We also demonstrate how our approach can be used to visualize the latent space of a StyleGAN v3 model and evaluate the variability of the generated images. The singularities of the Fisher metric in the two-dimensional section of the latent space are signs of a second-order phase transition, corresponding to a sharp change in the identity of the generated faces with a gradual change in the latent code. It is shown that the phase diagram of the generated images is mostly determined by the mapping network, so the synthesis network does not change the position of the phase boundaries. Potentially, this means that it is possible to extract semantically meaningful features by studying only the mapping network, since the most significant changes occur in directions orthogonal to the boundaries of the phase transition. In general, the reasons for the existence of the observed phase boundaries in StyleGAN models remain largely unclear, and this phenomenon requires further research.

\section*{Acknowledgements}

The authors acknowledge the use of Zhores HPC \cite{zacharov2019zhores} for obtaining the results presented in this paper. We are grateful to P. Krapivsky, V. Palyulin, A. Iakovlev and D. Egorov for interesting discussions. This work is supported by the RSF Grant No. 21-73-00176.

\begin{appendix}

\section{Learned posterior distributions for 2D Ising model}

\begin{figure}[H]
\vskip 0.2in
\begin{center}
\centerline{\includegraphics[width=0.75\columnwidth]{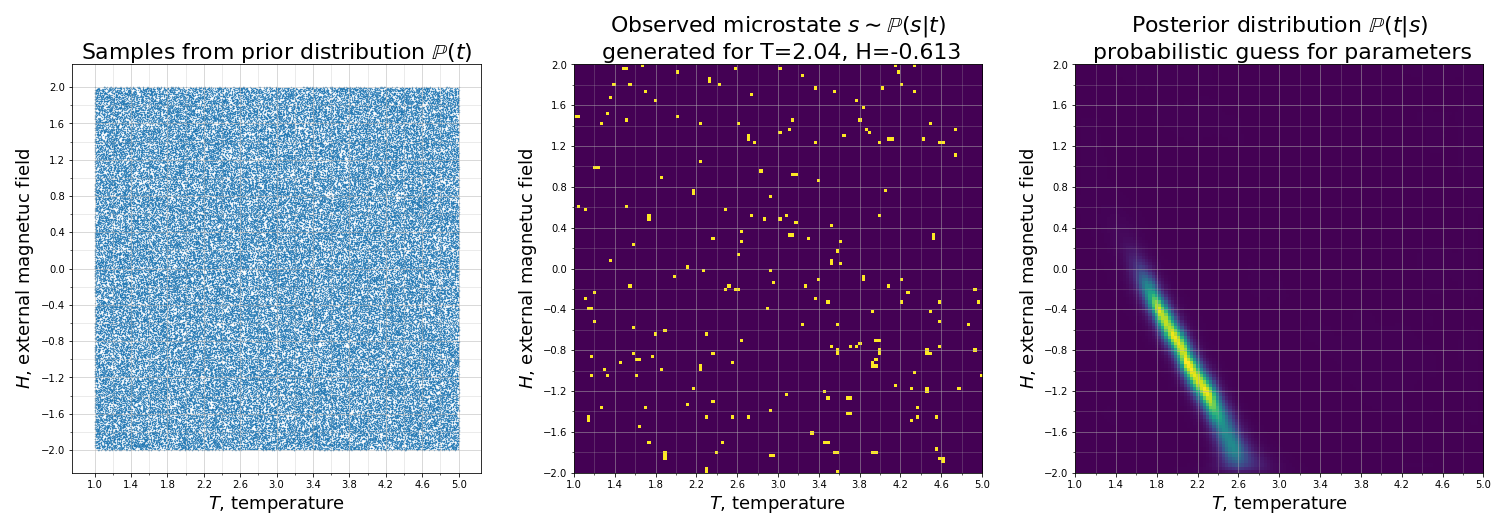}}
\caption{(left) Uniform prior distribution on the square $[T_{\min}, T_{\max}]\times[H_{\min}, H_{\max}] = [1, 5]\times[-2, 2]$. (center) Observed microstate of the Ising model generated for $T = 2.04, H = -0.613$. (right) Posterior distribution on the square $[T_{\min}, T_{\max}]\times[H_{\min}, H_{\max}]$ predicted by U$^{2}$Net.}
\label{tasep-free-energy}
\end{center}
\vskip -0.2in
\end{figure}

\begin{figure}[H]
\vskip 0.2in
\begin{center}
\centerline{\includegraphics[width=0.75\columnwidth]{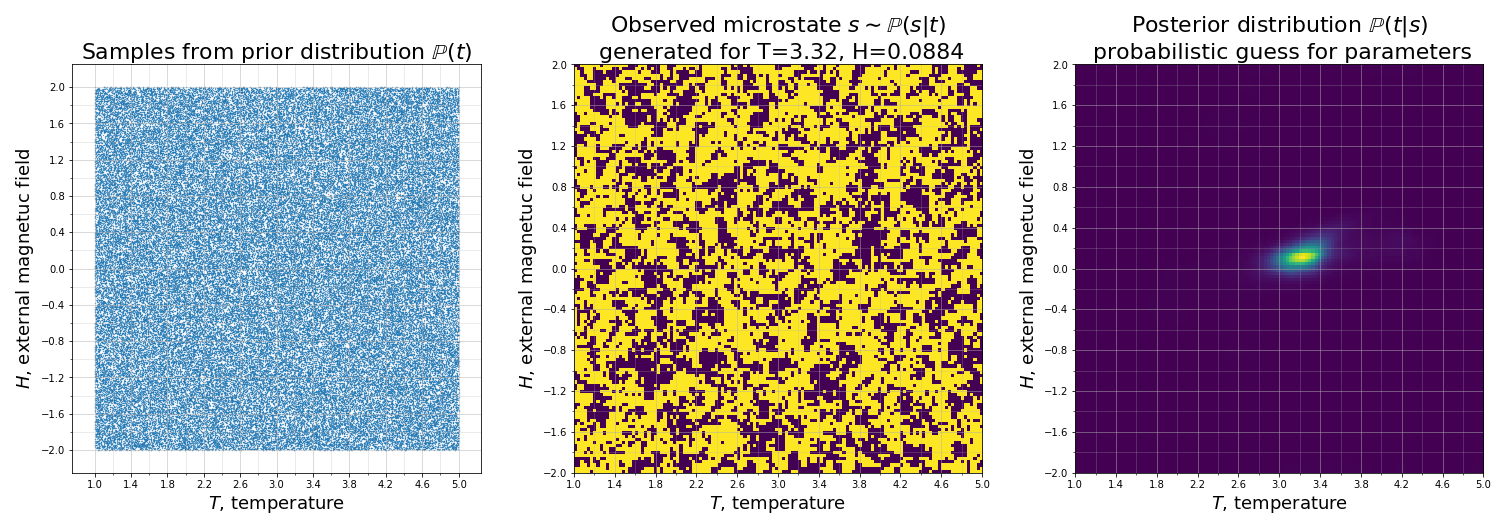}}
\caption{(left) Uniform prior distribution on the square $[T_{\min}, T_{\max}]\times[H_{\min}, H_{\max}] = [1, 5]\times[-2, 2]$. (center) Observed microstate of the Ising model generated for $T = 3.32, H = 0.0884$. (right) Posterior distribution on the square $[T_{\min}, T_{\max}]\times[H_{\min}, H_{\max}]$ predicted by U$^{2}$Net.}
\label{tasep-free-energy}
\end{center}
\vskip -0.2in
\end{figure}

\begin{figure}[H]
\vskip 0.2in
\begin{center}
\centerline{\includegraphics[width=0.75\columnwidth]{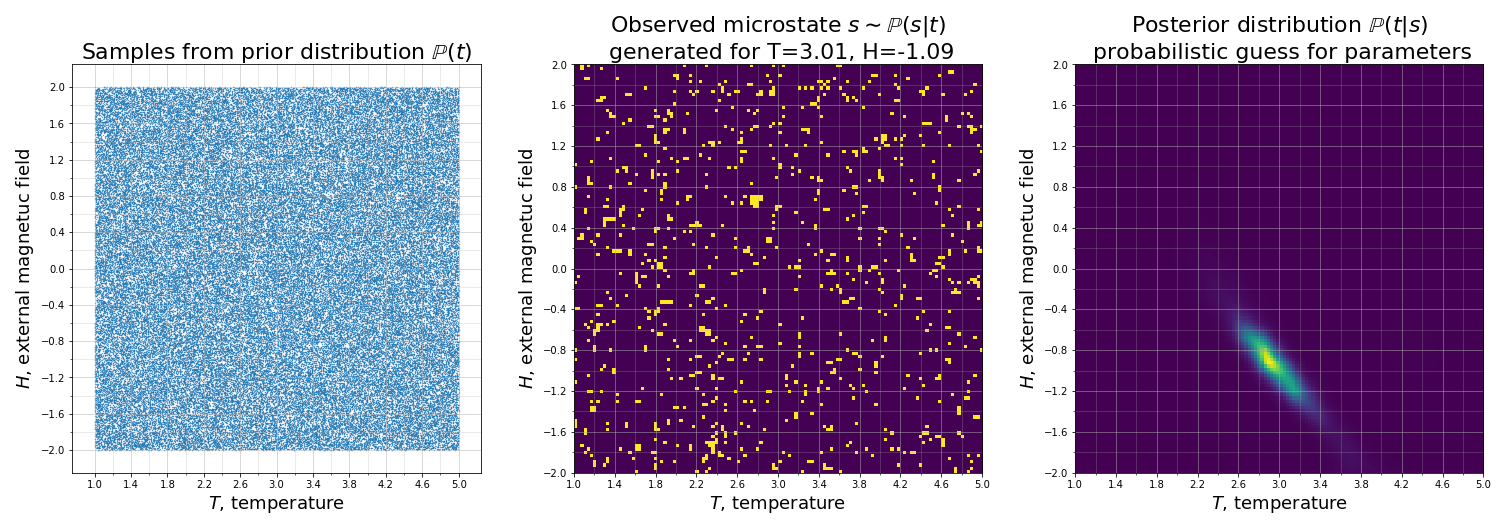}}
\caption{(left) Uniform prior distribution on the square $[T_{\min}, T_{\max}]\times[H_{\min}, H_{\max}] = [1, 5]\times[-2, 2]$. (center) Observed microstate of the Ising model generated for $T = 3.01, H = -1.09$. (right) Posterior distribution on the square $[T_{\min}, T_{\max}]\times[H_{\min}, H_{\max}]$ predicted by U$^{2}$Net.}
\label{tasep-free-energy}
\end{center}
\vskip -0.2in
\end{figure}

\section{StyleGAN v3 linear interpolation between three base faces}

\begin{figure}[H]
\vskip 0.2in
\begin{center}
\centerline{\includegraphics[width=\columnwidth]{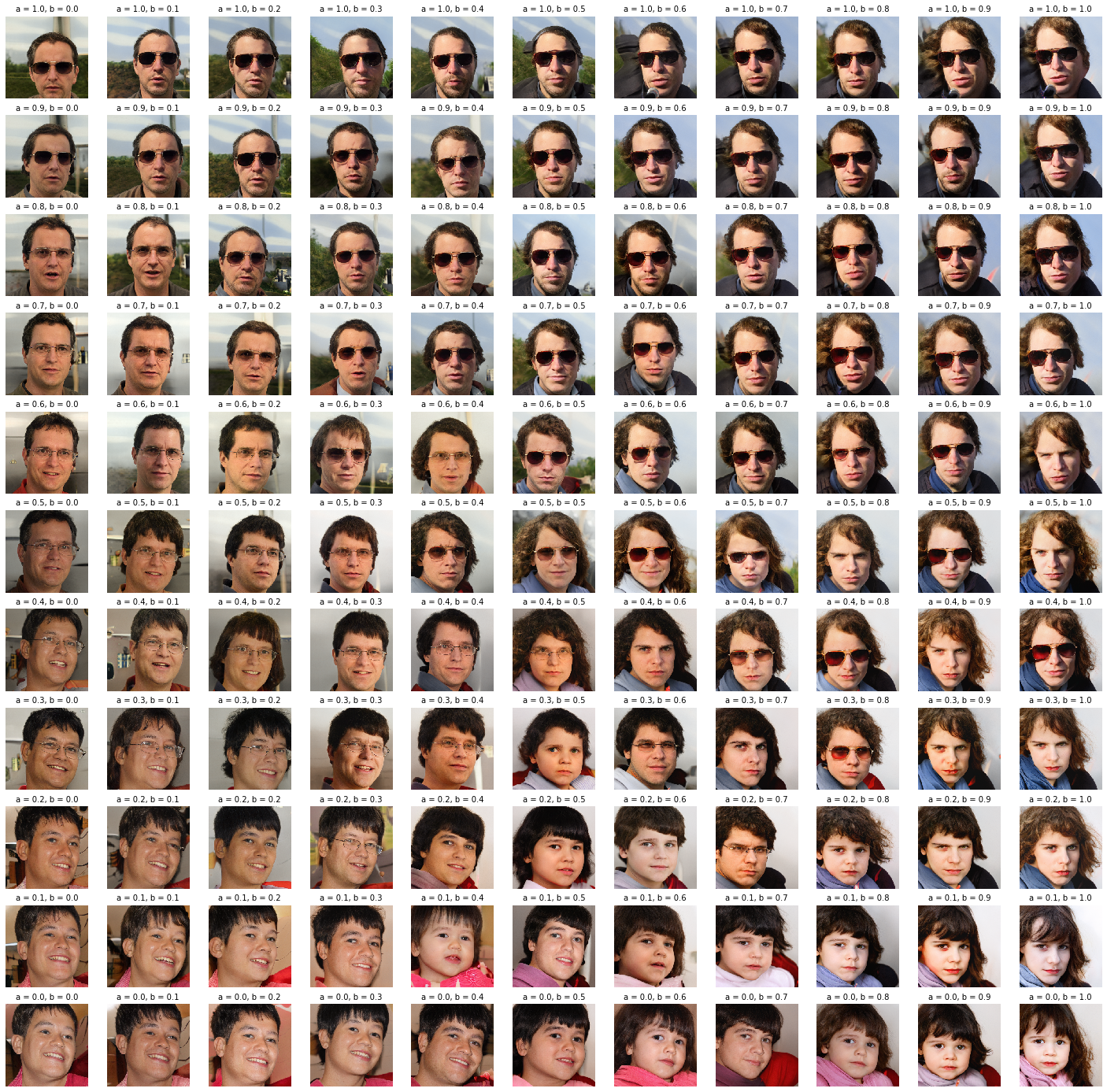}}
\caption{Landscape of generated faces on the plane of parameters $a, b$ of linear interpolation between three base faces. The discovered phase transitions occur between three macroscopic phases correspond to the  faces generated for $a=0, b=0$, $a+b \approx 1$ and $a=1, b=1$}
\label{generrated-faces-landscape-large}
\end{center}
\vskip -0.2in
\end{figure}

\section{Reconstructed Fisher metrics}

\begin{figure}[H]
\vskip 0.2in
\begin{center}
\centerline{\includegraphics[width=\columnwidth]{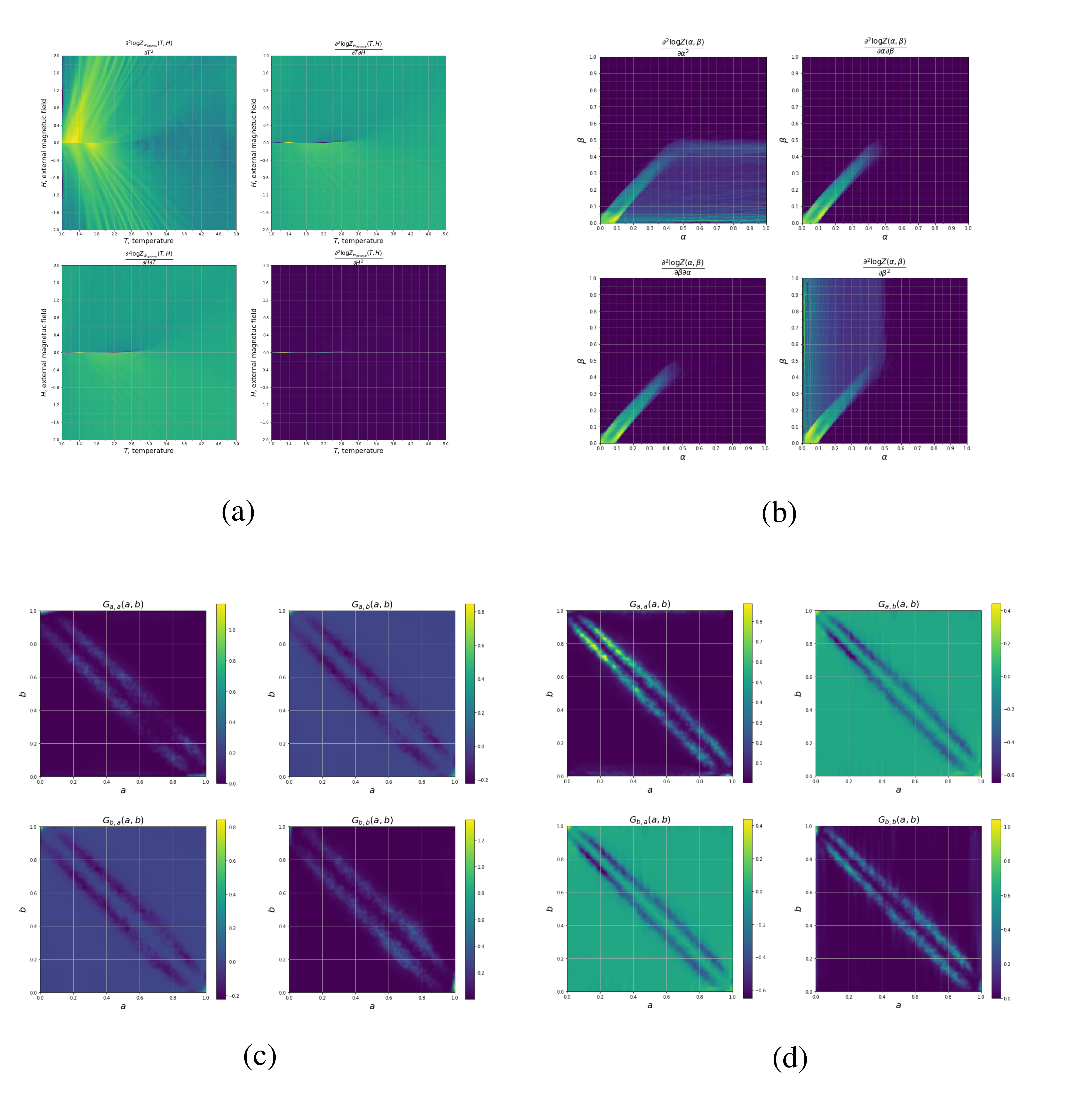}}
\caption{Components of the reconstructed Fisher metric. (a) 2D Ising model; (b) TASEP; (c) StyleGAN v3 trained on FFHQ with faces as microstates; (d) StyleGAN v3 trained on FFHQ with vectors from $\mathcal{W}$-space as microstates (only the mapping network is used).}
\label{reconstructed_fisher_metrics}
\end{center}
\vskip -0.2in
\end{figure}

\section{StyleGAN v3 latent space visualization}
\subsection{Density-based visualization of the posterior distribution on the space of macroscopic parameters}
\begin{figure}[H]
\vskip 0.2in
\begin{center}
\centerline{\includegraphics[width=\columnwidth]{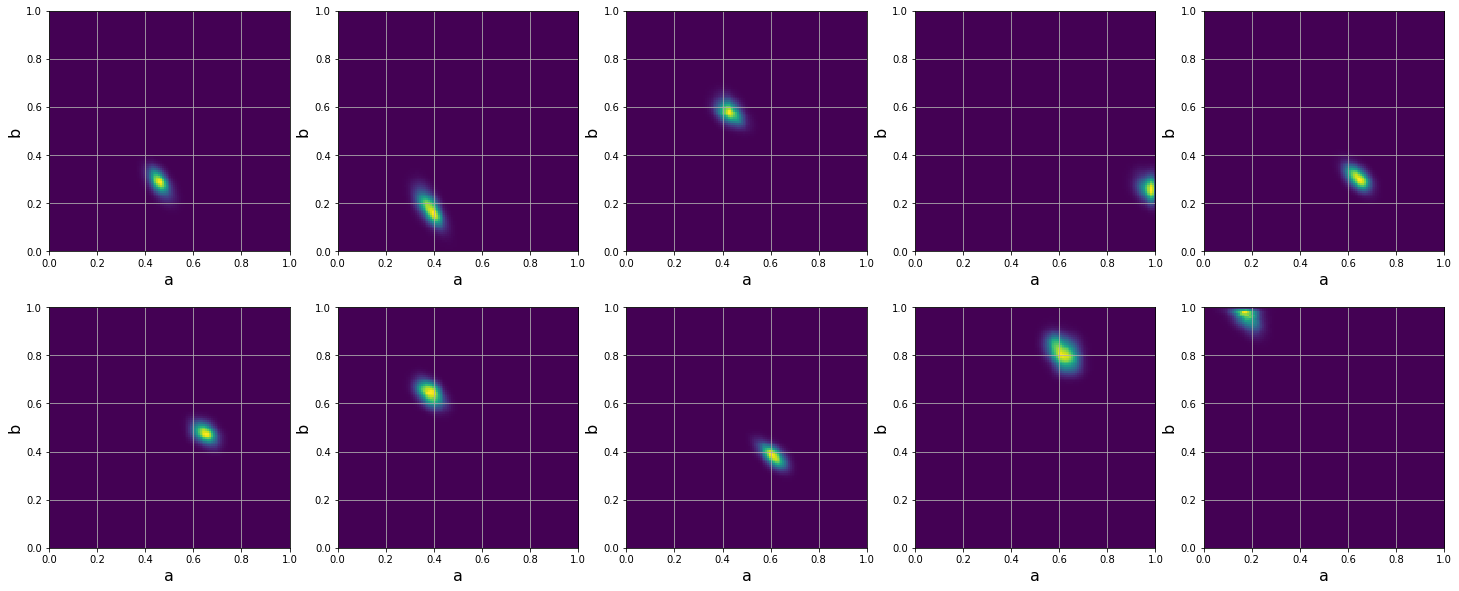}}
\caption{Posterior distributions on the space of parameters $a,b$ of the linear interpolation between three base faces. $M=1$ distributions per plot.  For every pixel maximum value of the density across all $M$ distributions was taken.}
\label{tasep-free-energy}
\end{center}
\vskip -0.2in
\end{figure}

\begin{figure}[H]
\vskip 0.2in
\begin{center}
\centerline{\includegraphics[width=\columnwidth]{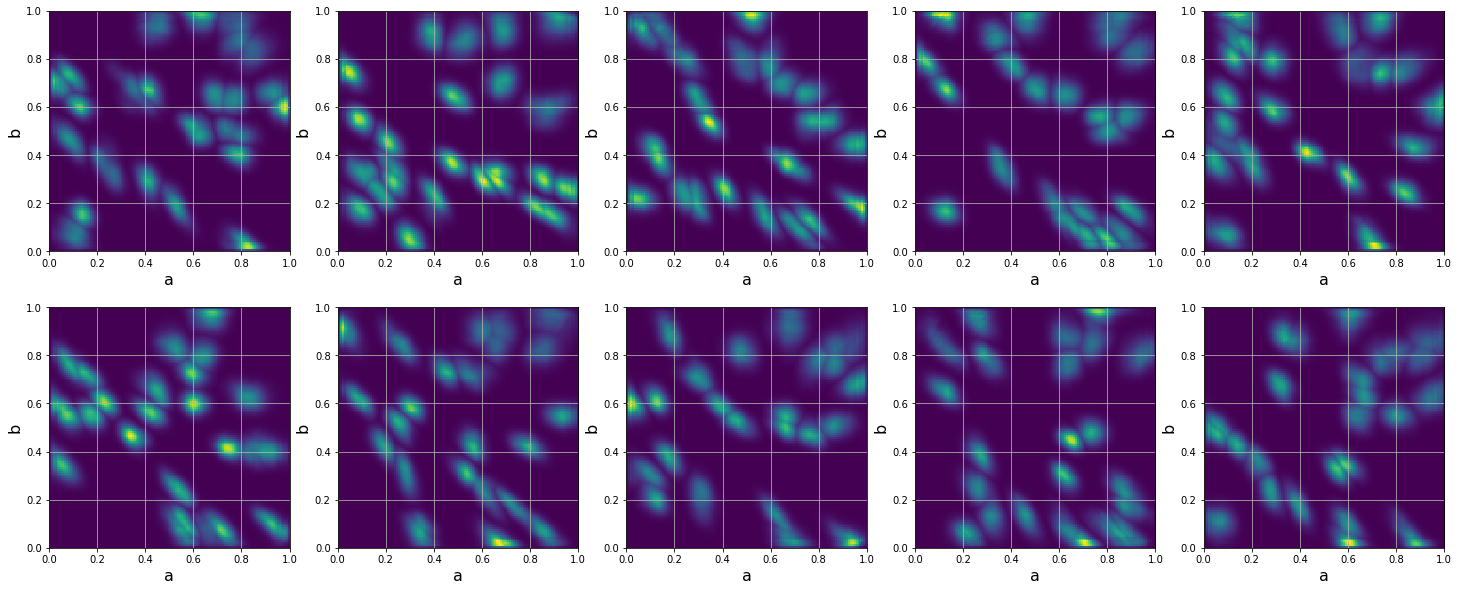}}
\caption{Posterior distributions on the space of parameters $a,b$ of the linear interpolation between three base faces. $M=25$ distributions per plot. For every pixel maximum value of the density across all  $M$ distributions was taken. }
\label{tasep-free-energy}
\end{center}
\vskip -0.2in
\end{figure}

\begin{figure}[H]
\vskip 0.2in
\begin{center}
\centerline{\includegraphics[width=\columnwidth]{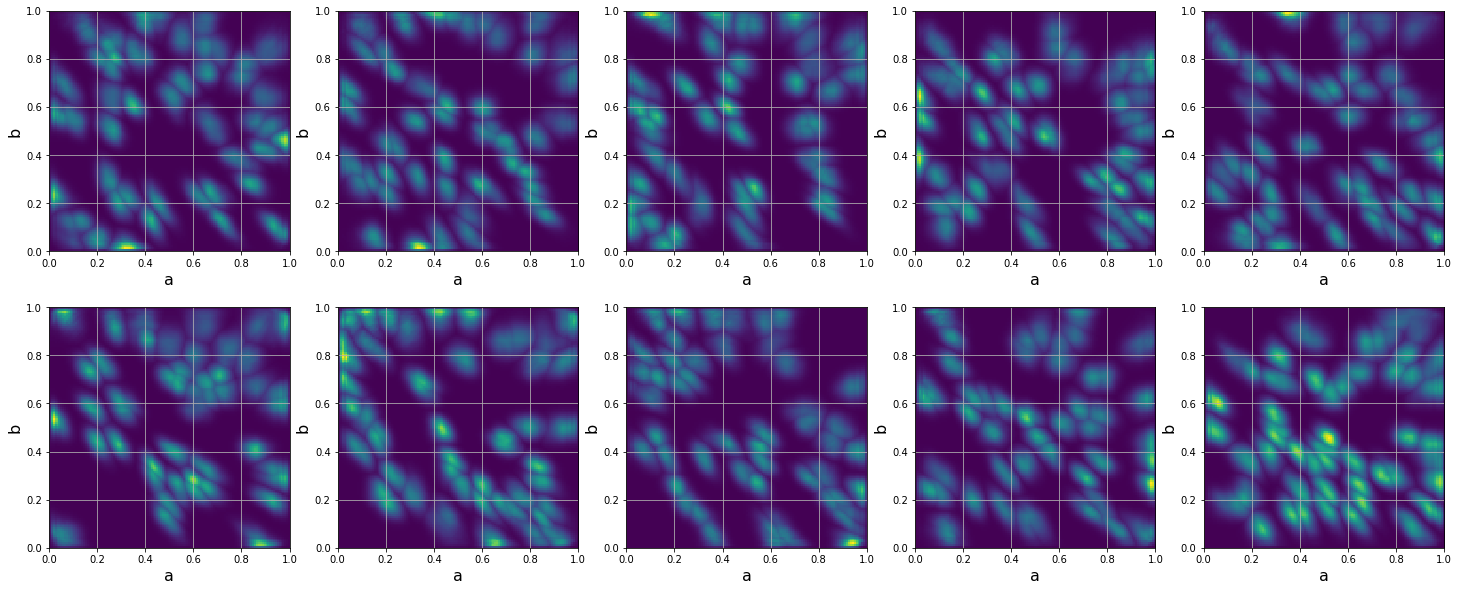}}
\caption{Posterior distribution on the space of parameters $a,b$ of the linear interpolation between three base faces. $M=50$ distributions per plot. For every pixel maximum value of the density across all  $M$ distributions was taken.}
\label{tasep-free-energy}
\end{center}
\vskip -0.2in
\end{figure}

\begin{figure}[H]
\vskip 0.2in
\begin{center}
\centerline{\includegraphics[width=\columnwidth]{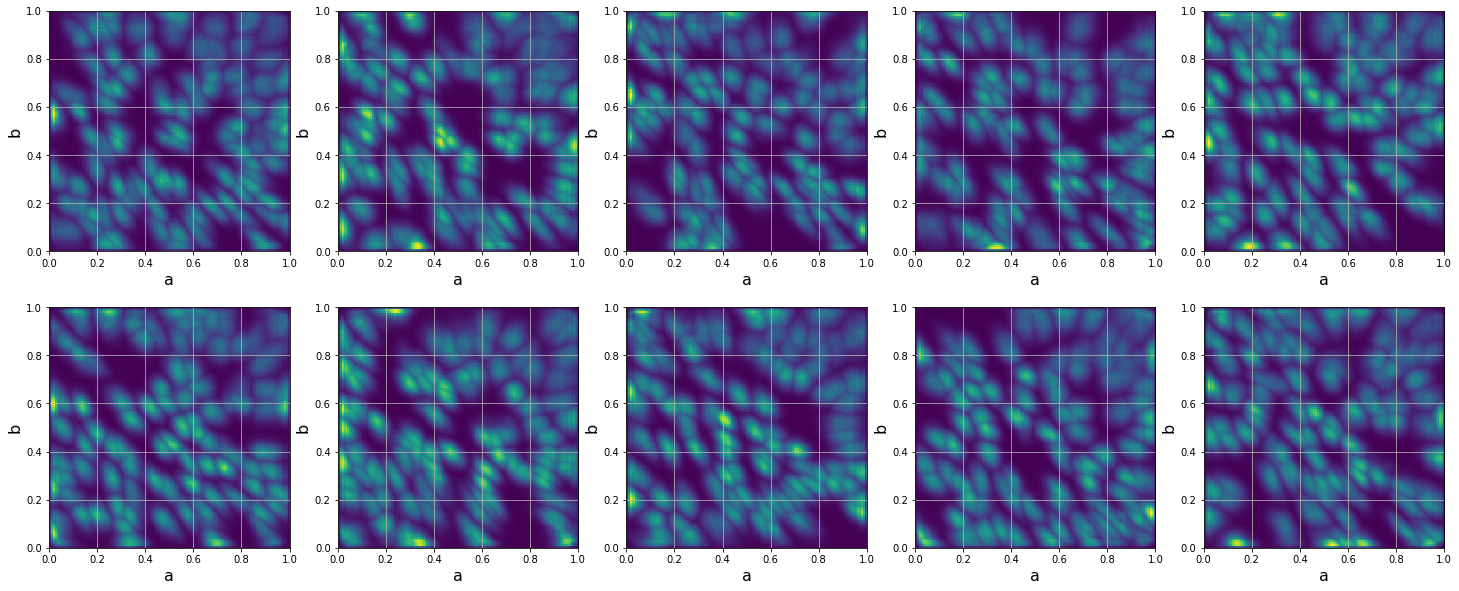}}
\caption{Posterior distribution on the space of parameters $a,b$ of the linear interpolation between three base faces. $M=100$ distributions per plot. For every pixel maximum value of the density across all  $M$ distributions was taken. }
\label{tasep-free-energy}
\end{center}
\vskip -0.2in
\end{figure}

\begin{figure}[H]
\vskip 0.2in
\begin{center}
\centerline{\includegraphics[width=\columnwidth]{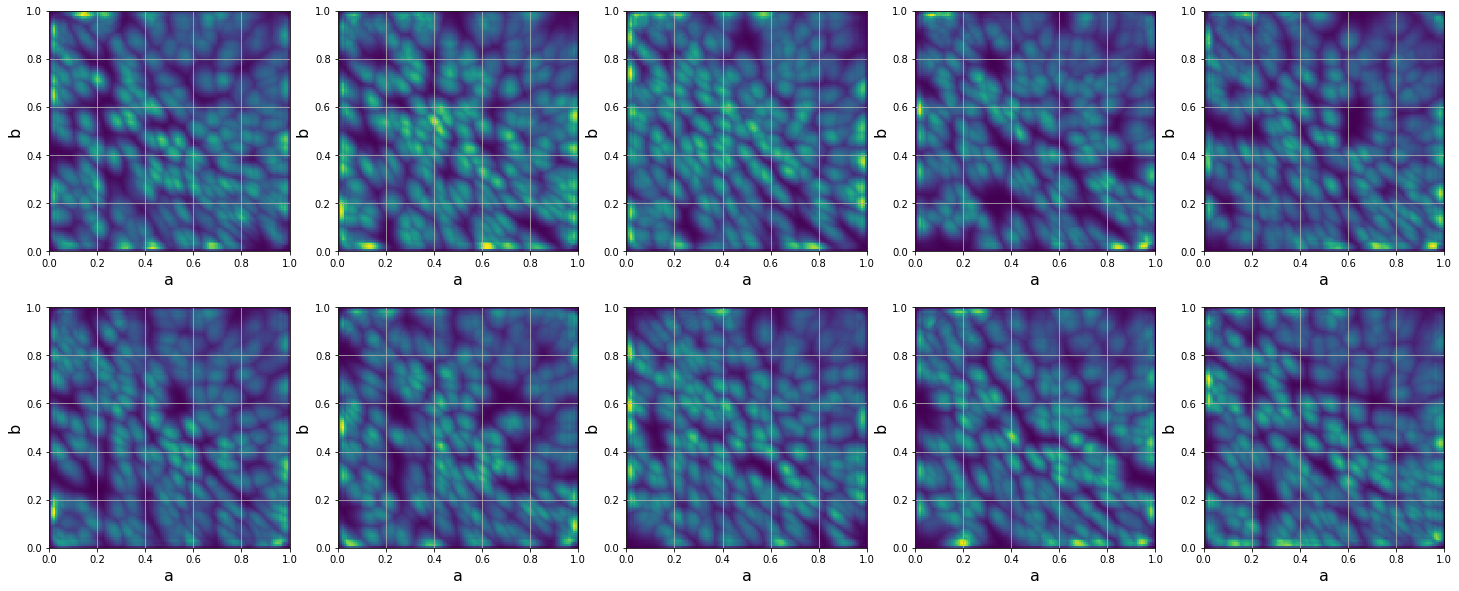}}
\caption{Posterior distribution the space of parameters $a,b$ of the linear interpolation between three base faces. $M=200$ distributions per plot.  For every pixel maximum value of the density across all distribution was taken.}
\label{tasep-free-energy}
\end{center}
\vskip -0.2in
\end{figure}

\subsection{Log-density-based visualization of the posterior distribution on the space of macroscopic parameters}

\begin{figure}[H]
\vskip 0.2in
\begin{center}
\centerline{\includegraphics[width=\columnwidth]{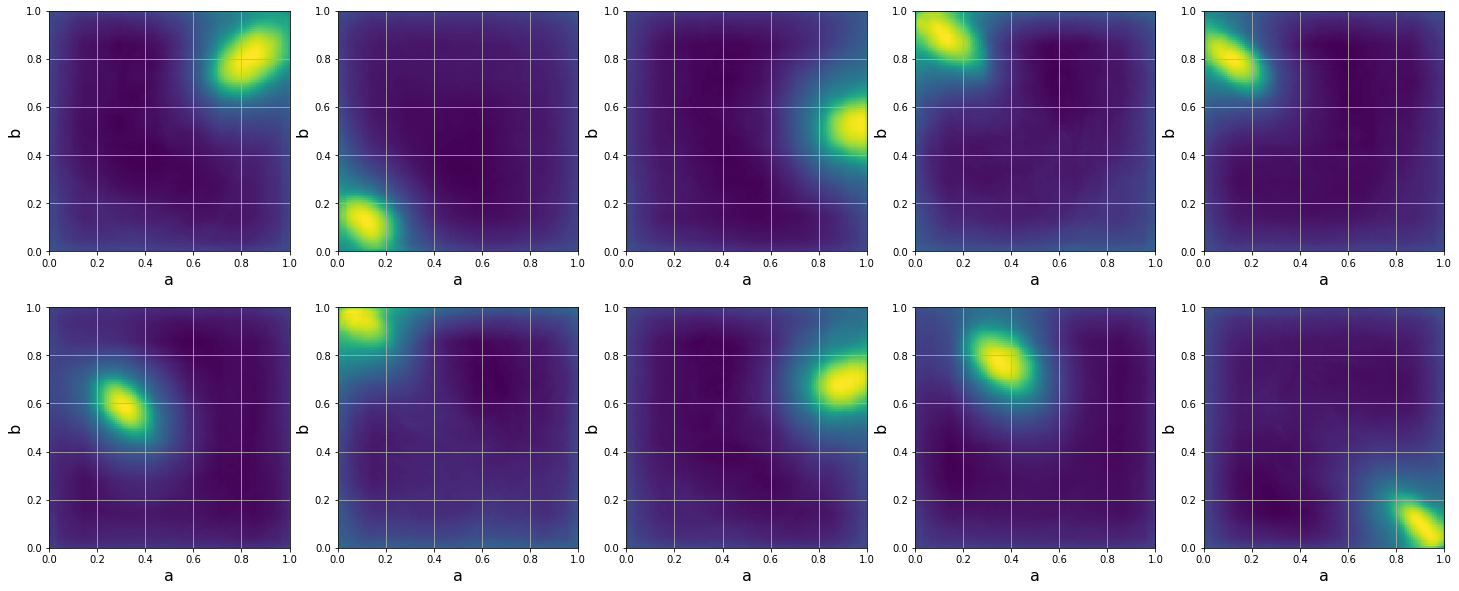}}
\caption{Logarithm of the posterior distributions on the space of parameters $a,b$ of the linear interpolation between three base faces. $M=1$ distributions per plot.  For every pixel maximum value of the log-density across all $M$ distributions was taken.}
\label{tasep-free-energy}
\end{center}
\vskip -0.2in
\end{figure}

\begin{figure}[H]
\vskip 0.2in
\begin{center}
\centerline{\includegraphics[width=\columnwidth]{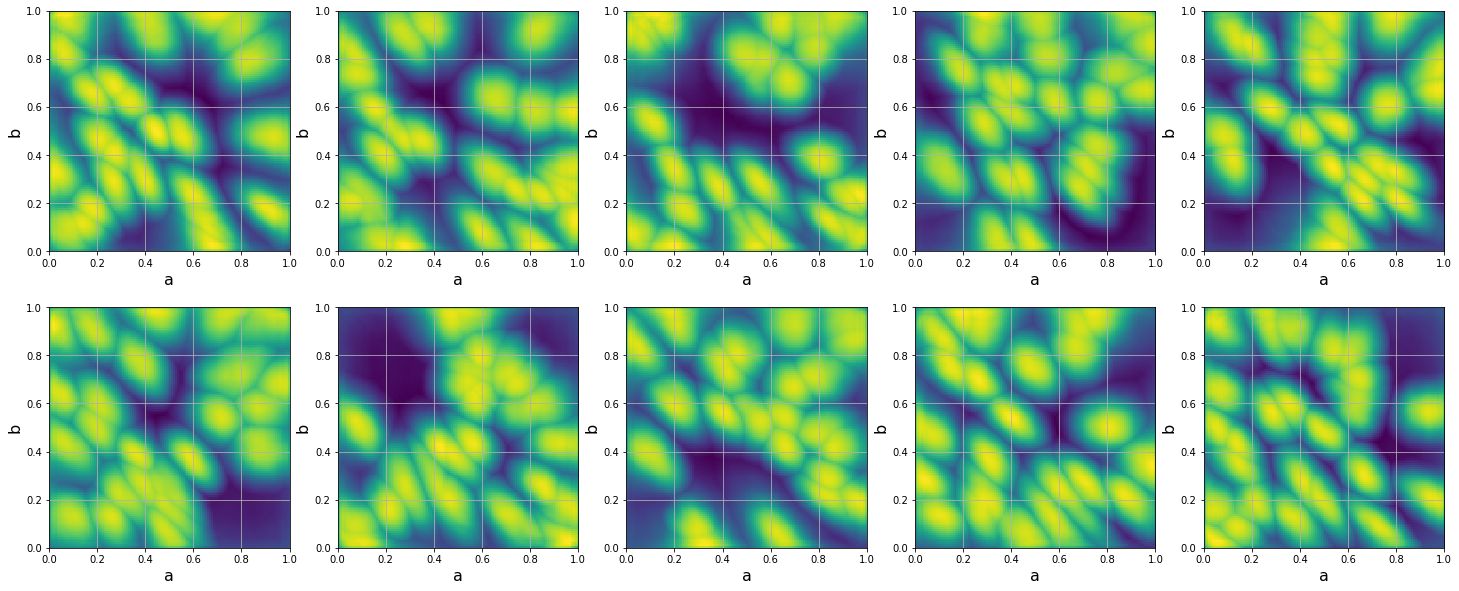}}
\caption{Logarithm of the posterior distributions on the space of parameters $a,b$ of the linear interpolation between three base faces. $M=25$ distributions per plot. For every pixel maximum value of the log-density across all  $M$ distributions was taken. }
\label{tasep-free-energy}
\end{center}
\vskip -0.2in
\end{figure}

\begin{figure}[H]
\vskip 0.2in
\begin{center}
\centerline{\includegraphics[width=\columnwidth]{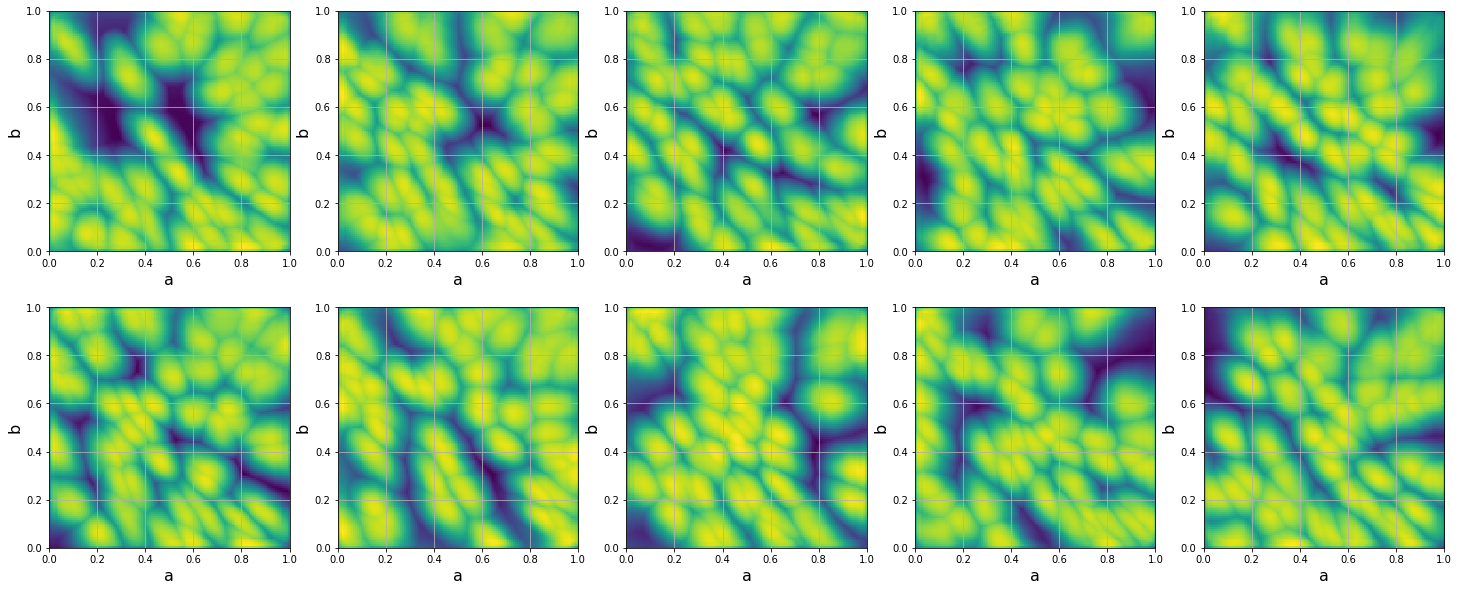}}
\caption{Logarithm of the posterior distribution on the space of parameters $a,b$ of the linear interpolation between three base faces. $M=50$ distributions per plot. For every pixel maximum value of the log-density across all  $M$ distributions was taken.}
\label{tasep-free-energy}
\end{center}
\vskip -0.2in
\end{figure}

\begin{figure}[H]
\vskip 0.2in
\begin{center}
\centerline{\includegraphics[width=\columnwidth]{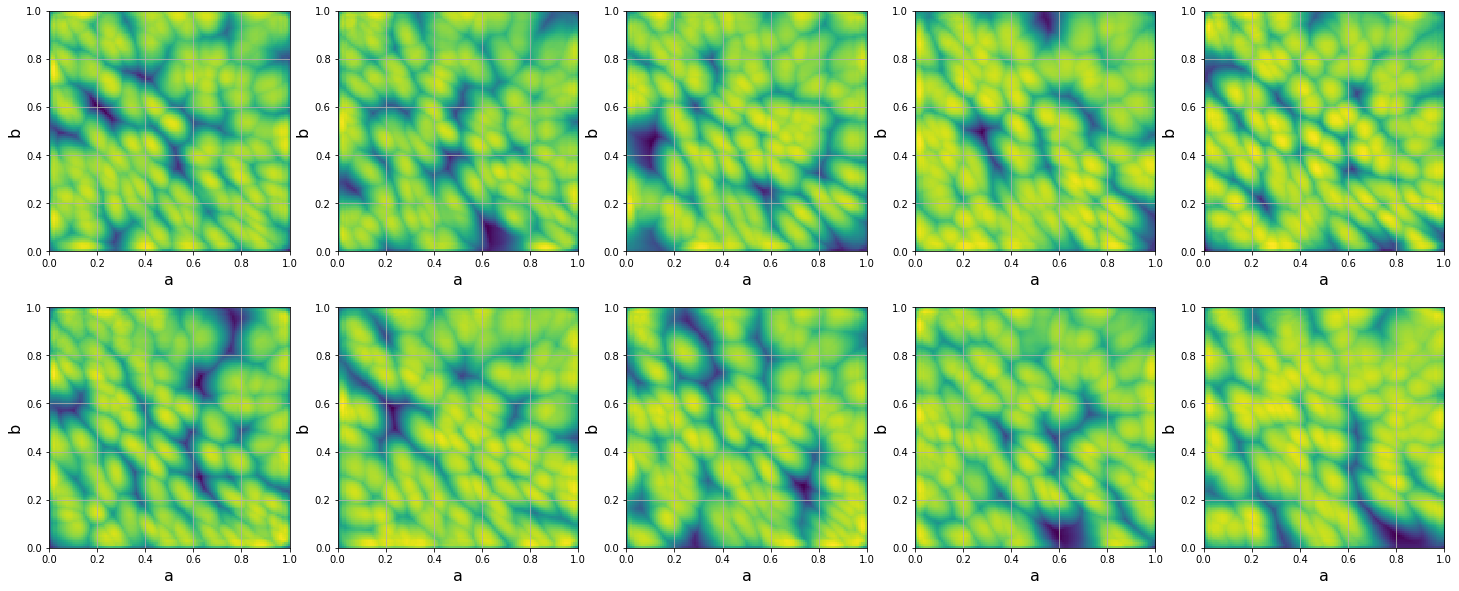}}
\caption{Logarithm of the posterior distribution on the space of parameters $a,b$ of the linear interpolation between three base faces. $M=100$ distributions per plot. For every pixel maximum value of the log-density across all  $M$ distributions was taken. }
\label{tasep-free-energy}
\end{center}
\vskip -0.2in
\end{figure}

\begin{figure}[H]
\vskip 0.2in
\begin{center}
\centerline{\includegraphics[width=\columnwidth]{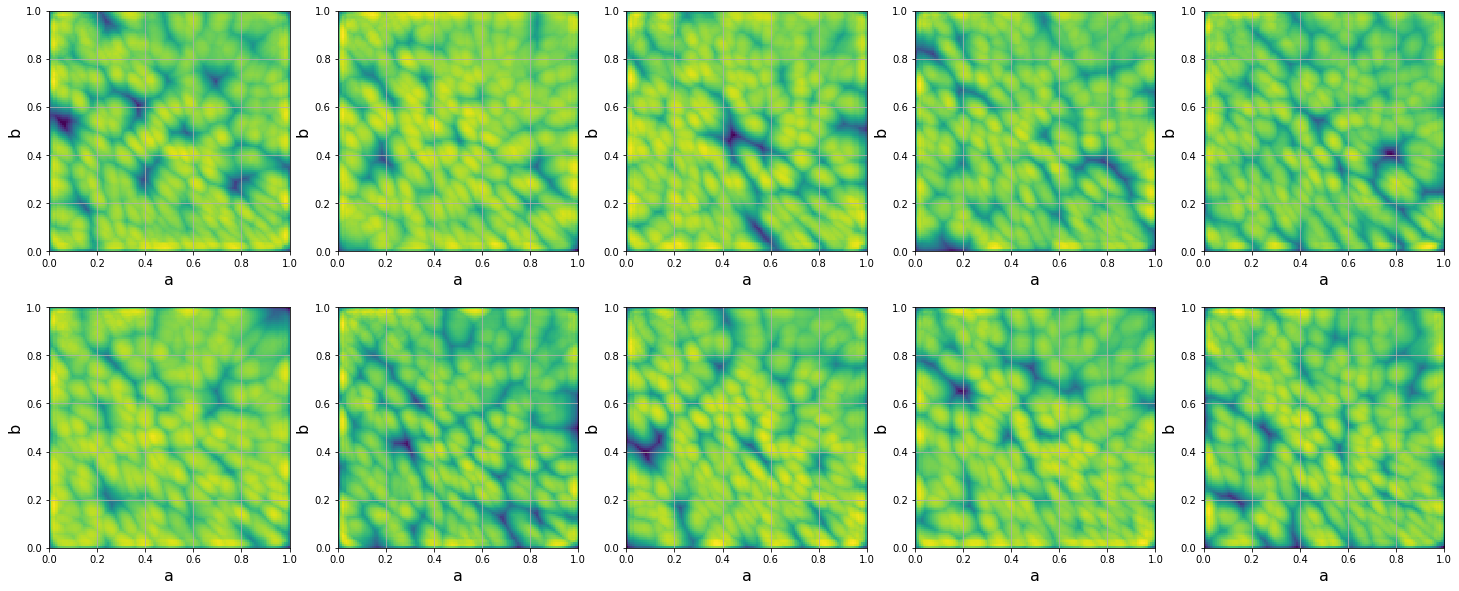}}
\caption{Logarithm of the posterior distribution the space of parameters $a,b$ of the linear interpolation between three base faces. $M=200$ distributions per plot.  For every pixel maximum value of the log-density across all distribution was taken.}
\label{tasep-free-energy}
\end{center}
\vskip -0.2in
\end{figure}
\end{appendix}

\end{document}